\documentclass[twocolumn]{article}
\pdfoutput=1
\usepackage[utf8]{inputenc}
\usepackage{amsmath,amsfonts,amssymb,amsthm}
\usepackage{graphicx}
\usepackage[caption=false,font=footnotesize]{subfig}
\usepackage{epstopdf}
\usepackage{color}
\usepackage{etoolbox}
\usepackage[hidelinks]{hyperref}
\graphicspath{ {./figures/} }

\hypersetup{pdftitle={Fusing Loop and GPS Probe Measurements to Estimate Freeway Density}}

\newtoggle{eprint}
\toggletrue{eprint}

\iftoggle{eprint}{\usepackage[margin=.75in,letterpaper]{geometry}}{}

\title{Fusing Loop and GPS Probe Measurements to Estimate Freeway Density}
\author{Matthew Wright\iftoggle{eprint}{}{,~\IEEEmembership{Student Member,~IEEE},} and Roberto Horowitz\iftoggle{eprint}{}{,~\IEEEmembership{Senior Member,~IEEE}}
		}

\begin{document}
\maketitle

\begin{abstract}
In an age of ever-increasing penetration of GPS-enabled mobile devices, the potential of real-time ``probe'' location information for estimating the state of transportation networks is receiving increasing attention. Much work has been done on using probe data to estimate the current speed of vehicle traffic (or equivalently, trip travel time).
While travel times are useful to individual drivers, the state variable for a large class of traffic models and control algorithms is vehicle density.
Our goal is to use probe data to supplement traditional, fixed-location loop detector data for density estimation.
To this end, we derive a method based on Rao-Blackwellized particle filters, a sequential Monte Carlo scheme.
We present a simulation where we obtain a 30\% reduction in density mean absolute percentage error from fusing loop and probe data, vs. using loop data alone.
We also present results using real data from a 19-mile freeway section in Los Angeles, California, where we obtain a 31\% reduction.
In addition, our method's estimate when using only the real-world probe data, and no loop data, outperformed the estimate produced when only loop data were used (an 18\% reduction).
These results demonstrate that probe data can be used for traffic density estimation.
\end{abstract}
\iftoggle{eprint}{}{
\begin{IEEEkeywords}
 Filtering algorithms, Global Positioning System (GPS), Hidden Markov model, Particle filters, Road transportation, State estimation
\end{IEEEkeywords}
}
\newtheorem{assumption}{Assumption}

\section{Introduction}

\label{sec:introduction}
\iftoggle{eprint}{Modeling}{\IEEEPARstart{M}{odeling}} the flow patterns of traffic throughout a road network is a key area of concern for traffic engineers and civil planners. When compared to many other objects studied in a systems context, road networks exhibit high levels of nonlinear phenomena in congestion shockwaves~\cite{blandin2012sequential}, have relatively low levels of sensor penetration~\cite{patire2015howmuchdata} (typically fixed sensors that measure vehicle flows), and what sensor measurements are available may exhibit high degrees of bias or noise~\cite{chen2003errors}. The sparsity and inaccuracy of detection makes estimating the spatiotemporal state of traffic flows difficult. 

Modeling the evolution of traffic flows over time, on the other hand, has been possible to achieve with high accuracy using partial differential equations (PDEs) based on fluid flows~\cite{lighthill1955kinematic, richards1956shock}. These \emph{macroscopic flow models}, particularly time-space discretizations suitable for computer simulation, have been widely adopted for capturing and forecasting traffic patterns on an aggregate scale.
A natural application of these models is in algorithms for estimating the state of unobserved areas of the road or debiasing erroneous measurements. State estimation methods based on physical traffic models have advantages over those based purely on statistical models. Physical models allow for estimation of conditions at uninstrumented locations, as well as prediction of network behavior in response to previously-unobserved conditions (for example, increased demands on a road network in response to a special event). Extrapolation of this sort with a statistical model is of course risky. Instead, the physical model is used to generate a state trajectory that best fits the sensor data, but the nonlinearity of the traffic PDEs makes this a hard problem to solve. Since the traffic PDEs are nonlinear, finding a solution that best matches he observed data across all space and time is not easy without considerable relaxations \cite{claudel_convex_2011}. More common is to treat the problem in a filtering context, where estimates are propagated forward in time through traffic models and updated with information from measurements. Nonlinear filters used for traffic state estimation in the literature include the Extended Kalman Filter~\cite{wang2005EKF} and the Sequential Monte Carlo-based Mixture Kalman~\cite{sun2004mixturekalmanfilter}, Ensemble Kalman~\cite{work2009trafficmodel} and particle filters~\cite{mihaylova2007PF}.

Traditional sensors for filtering on road networks, such as buried inductive loop detectors or video cameras, have been fixed in location. Recently, though, there has been great interest towards augmenting these fixed data with data collected by other parties as part of the current explosion of sensor availability and data collection~\cite{work2009trafficmodel}.
Transportation authorities have been eager to leverage these new data as both a low-cost alternative to increasing penetration of detection and to extend detection to areas for which installation is economically infeasible~\cite{patire2015howmuchdata}. 
Works such as~\cite{work2009trafficmodel} have successfully used measurements of individual vehicles' velocity collected from passengers' GPS-enabled mobile devices, or \emph{probe} data, for the filtering problem.
This supplementation is principally referred to as \emph{data fusion} in the transportation literature~\cite{patire2015howmuchdata}. 

The method of~\cite{work2009trafficmodel} is perhaps the most popular data fusion method, with recent studies examining the marginal gains for varying data quantities in simulation~\cite{bucknell2014tradeoff} and deployment at several real-world sites~\cite{allstrom2012MMStockholm, patire2015howmuchdata}. These results and others have shown the availability of probe data makes it useful in estimating the traffic state across a large freeway corridor, on the order of tens of miles.

Unfortunately, the state estimation method in~\cite{work2009trafficmodel} is only intended to estimate the state of traffic's mean velocity (which is then easily converted into travel times, a common metric of road performance), whereas traffic control algorithms tend to rely on estimates of the density of vehicles~\cite{papageorgiou2003controlReview}. As we will explain below, estimating density directly from velocity measurements is difficult. A method for supplementing data from fixed detection infrastructure with probe data while retaining the model-based filtering approach has been elusive, but we show that reconsidering the problem in a probabilistic view simplifies the mathematics, and further develop an approach based on Rao-Blackwellization of particle filters as a solution.

The remainder of this article is structured as follows. Sections \ref{sec:macroscopicModels} and \ref{sec:filteringIntro} provide a brief review of macroscopic flow models, describes the specific model used in this work, and discusses the filtering problem in a general setting. In Section \ref{sec:filteringCTM} we discuss the specific case of filtering on macroscopic flow models from a probabilistic perspective. We also provide a motivation for ensemble methods for the traffic filtering problem and give a formulation of a general particle filtering algorithm for use with any macroscopic flow model. Section \ref{sec:rbpf} extends the general model developed in Section \ref{sec:filteringCTM} to the specific problem of real-time assimilation of probe velocity measurements to augment traffic density estimates, and Section \ref{sec:experiments} provides some numerical experiments demonstrating the applicability of velocity measurements towards estimating traffic density. Finally, Section \ref{sec:conclusion} provides some closing thoughts and discusses several immediate uninvestigated problems.

\section{Traffic flow models and filtering}
\label{sec:flowmodelsFiltering}
\subsection{Macroscopic flow models}
\label{sec:macroscopicModels}

This article uses a macroscopic model of vehicular traffic. This type of model abstracts traffic flows along a road as a 
fluid flow.
In other words, while traffic flows are actually made of many
individual vehicles acting independently, the dynamics of flows 
along a long, straight road at large scale may be modeled as evolving due to a
one-dimensional continuity equation of the form \cite{lighthill1955kinematic,richards1956shock}
\begin{equation}
\frac{\partial \rho(z,t)}{ \partial t} + \frac{ \partial q(z,t)}{\partial z} = 0, \label{eq:LWR}
\end{equation}
where $\rho$ is the density of vehicles at lineal location $z$ at time $t$, and
$q(\cdot)$ is some flux function.
This construction is called the Lighthill-Whitham-Richards (LWR) model of 
traffic flow.
Today, simulation methods
based on LWR and other macroscopic descriptions are used to model 
traffic on freeways so that public authorities may estimate congestion for the purposes of
traffic control and infrastructure planning~\cite{kurvar15}.

The LWR model of traffic \eqref{eq:LWR} is often said to be stated in \emph{Eulerian} coordinates.
This designation refers to the Eulerian characterization of a fluid flow field, in which the state of the flow field is parameterized by space $z$ and time $t$.
In contrast is the \emph{Lagrangian} characterization of a flow field, which tracks the position of individual fluid elements.
In the context of vehicle traffic, the fluid elements are vehicles, and the flow field is parameterized in terms of individual vehicle number and time \cite{leclercq_lagrangian_2007}.

Along these lines, traffic data taken from fixed sensors are often called Eulerian data, and data that describe individual vehicles (i.e., our probe data) are called Lagrangian data.

A common type of Eulerian sensor is the buried inductive loop detector, which detects the presence of a vehicle as it passes.
Later in this article, we will use loop data from the California Performance Measurement System (PeMS) \cite{pems_web}. The PeMS dataset contains measurements from single-loop and double-loop detectors; double-loop detectors measure vehicle flows and speeds directly (from which vehicle densities are inferred), while the single-loop detectors measure vehicle flows and estimate speeds and densities using the PeMS ``g-factor'' algorithm~\cite{jia2001gfactor}.

\begin{figure}
	\centering
	\def\svgwidth{0.75\columnwidth}
	\iftoggle{eprint}{
	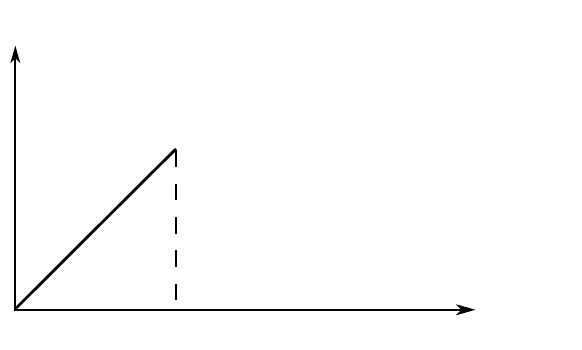}{
	\input{./figures/fundamentalDiagram.pdf_tex}}
	\caption{Schematic illustration of a Fundamental Diagram. The horizontal axis $\rho$ is vehicle density (typically veh/m) and the vertical axis $q$ is vehicle flow (typically veh/s).}
	\label{fig:FD}
\end{figure}

In this paper, we use as our traffic model the Cell Transmission Model (CTM) originally due to Daganzo \cite{daganzo1994ctm} with some modifications.
The CTM is a finite-volume approximation of \eqref{eq:LWR} that breaks roads into small discrete segments with homogeneous density.
We refer to these road segments as links, and the locations where they are joined as nodes.
A key component of the CTM is the \emph{fundamental diagram}, which describes the flux function $q(\cdot)$ as a function of a link's density.
Here, we use a triangular fundamental diagram of the form
\begin{equation}
	q_\ell = \min \left( v_{f,\ell} \cdot \rho_\ell, \; w_\ell \cdot \left( \rho_{j,\ell} - \rho \right) \right), \label{eq:FD}
\end{equation}
where $\ell$ is an index that denotes a particular link, $\rho_\ell$ is the density of a link, $q_\ell$ is the flow on link $\ell$, $v_{f,\ell}$ is the freeflow speed of the link, $w$ is the link's congestion wave speed, or the maximum speed at which shockwaves move backward through the link, and $\rho_{j,\ell}$ is the jam density, or the maximum possible density of a link, at which no point no more vehicles may be accommodated (Fig.~\ref{fig:FD}).

The fundamental diagram captures all parameters of a link's behavior. In applications to real road networks, these parameters are fit to data~\cite{munoz2004ctmcalibration}. A key quantity of interest that can be computed from the fundamental diagram is the average velocity of the link, which is simply
\begin{align}
  v\left( \rho \right) = \frac{q\left( \rho \right)}{\rho} = \frac{\min \left(R\left( \rho \right), S \left( \rho \right) \right)}{\rho}. \label{eq:velocity}
\end{align}

In the CTM, the flow between an upstream link $\ell$ and a downstream link $\ell+1$ is a function of both $\ell$ and $\ell+1$.
See \cite{daganzo1994ctm} for details on the situation where a link has only one downstream link, and that downstream link has only the one upstream link.

More complex situations occur when links have more than one incoming or outgoing link. In the context of this paper, which focuses on freeways, the only relevant situations are merges of an onramp and the freeway and diverges to an offramp in the freeway. For onramps and offramps, in this paper we use a simple merge model proposed by~\cite{muralidharan2009lnctm}.
In the \cite{muralidharan2009lnctm} onramp merge model, the flows exiting the upstream mainline link and the onramp link are proportional to proportional to the links' demand, which is itself a function of the links' density.

Driver behavior at offramp diverges is parameterized by a \emph{split ratio} coefficient $\beta_\ell \in [0,1]$, which is the portion of vehicles in link $\ell$ that wish to enter the offramp.
The offramps are assumed to accept all vehicles that wish to enter them (that is, they are assumed to never be in congestion).
For more information as to the particularities of this offramp mode, we refer again to \cite{muralidharan2009lnctm}.

The original CTM of Daganzo has been characterized as a \emph{first-order} model, as the dynamic equation of state is a function of only one variable, $\rho_\ell$. Various authors have proposed alternative parameterizations of the sending and receiving functions, or even higher-order parameterizations of a link's state, where the one-step update calculation is a function of the density and one or more other quantities. We discuss a few second-order models used for filtering in Section~\ref{sec:filteringCTM}.

We have thus far described a deterministic model for traffic flow, but in a filtering context a stochastic model is required. The large number of parameters in the CTM lead to a rich capability to introduce stochasticity into the model. In particular, authors have proposed stochastic models that include uncertainty in the fundamental diagram parameters, upstream (boundary condition) demands, or driver behavior at diverges. For the simulations later in this paper, we treat the onramp and offramp parameters as stochastic. The onramp flows are contaminated with additive white Gaussian noise (similar to in e.g. \cite{wang2005EKF}), and the split ratios $\beta$ are taken as independent-across-time beta-distributed random variables \cite{wright2015splits}. Both distributions were fit to data. Note that the non-correlation across time is a modeling assumption, and a more realistic model would consider a nontrivial autocorrelation of these time-varying random variables.

\subsection{Probabilistic Systems and Filtering}
\label{sec:filteringIntro}
The probability calculations in this paper are presented in terms of probability density functions (PDFs); measure theoretic rigor is omitted for accessibility. We adopt a nomenclature of probabilistic state-space systems to be consistent with much of the literature on particle filtering. In particular, let $x_t$ be the (unobserved) state of the system of interest (in our case, the vector of link densities) at time $t$ and $y_t$ be the observation of the system at the same time. The variables evolve over time through discrete-time stochastic state and output equations, denoted $\mathcal{F}_\theta(\cdot)$ and $\mathcal{G}_\theta(\cdot)$ respectively:
\begin{align}
\label{eq:fAndg}
\begin{split}
  x_t &= \mathcal{F}_\theta \left(x_{t-1} \right) \\
  y_t &= \mathcal{G}_\theta \left(x_t \right),
\end{split}
\end{align}
with $\theta$ a parameter vector describing the randomness or process/measurement noise of $\mathcal{F}$ and $\mathcal{G}$. $\mathcal{F}_\theta(\cdot)$ is a shorthand for the one-step CTM update described above. The notation~\eqref{eq:fAndg} is equivalent to
\begin{align}
  \begin{split}
    X_t | \left(X_{t-1} = x_{t-1} \right) 
	    &\sim f_{\theta, X_t|X_{t-1} = x_{t-1}} \left(x_t|x_{t-1} \right) \\
    Y_t | \left(X_t = x_t \right) 
	    &\sim g_{\theta, Y_t|X_t = x_t} \left(y_t|x_t \right), \label{eq:xAndYAsRVs}
  \end{split}
\end{align}
where $X_t$ $(Y_t)$ denotes a random variable and $x_t$ $(y_t)$ the value of a particular realization. The functions $f( \cdot )$ and $g( \cdot )$ are the PDFs induced by $\mathcal{F}_\theta( \cdot)$ and $\mathcal{G}_\theta(\cdot)$, respectively. The initial condition of the system, $x_0$, is assumed fixed or distributed with some known density $p_{\theta,X_0}(x_0)$. More precisely, $f_{\theta, X_t | X_{t-1} = x_{t-1}} (x_t|x_{t-1})$ is a Markov transition kernel with a distribution on the random variable $X_t|(X_{t-1}=x_{t-1})$, and $g_{\theta, Y_t|X_t = x_t}(y_t|x_t)$ is a typical observation PDF. Of importance is that \eqref{eq:xAndYAsRVs} establishes that the one-step update and measurement equations may be used as probability densities for the random variables $X_t | (X_{t-1}=x_{t-1})$ and $ Y_t | (X_t=x_t)$.

For the remainder of this paper, we will use two notational shorthands. First, we will drop the subscript of the random variable in writing PDFs, e.g. $f_{\theta, X_t | X_{t-1} = x_{t-1}} (x_t|x_{t-1})$ will be written as $f_\theta (x_t|x_{t-1})$, and the conditional random variables will have the value conditioned on omitted, e.g. $X_t | (X_{t-1} = x_{t-1})$ will be written as $X_t | X_{t-1}$.

A framework of this form is often referred to as a \emph{Hidden Markov Model} (HMM), after the Markov structure of the unobserved variable $x$, or in the specific case where $x$ and $y$ are real-valued vectors, a state-space model. 
The central problem is inference on the unobserved process $X$ using the information from the observed process $Y$, i.e. the formulation of conditional PDFs of the form $p_\theta( x_{t_1}|y_{t_2} )$ for some timesteps $t_1$ and $t_2$. In the present work, we are particularly interested in the filtering problem, which seeks at time $t$ the PDF $p_\theta\left( x_t | y_1, y_2, \dots y_{t-1}, y_t \right)$, or the PDF of the state at the current time conditioned on all observations received up until the current time. As an additional notational shorthand, let us denote as $y_T$ the collection of observations from the initial time to time $t$ inclusive, i.e. $y_T$ = $\{y_1, y_2, \dots, y_{t-1}, y_t\}$. Similarly, let $y_{T-1} = \{y_1, y_2, \dots, y_{t-2}, y_{t-1}\}$.

The filtering problem is typically solved in recursive one-step updates. An intuitive explanation of the recursive filtering scheme in a probabilistic sense begins by noting that due to the assumed HMM structure, we have:
\begin{align}
  \begin{split}
    p_\theta \left(x_t|x_{1,\dots,t-1},y_{T-1} \right) &= f_\theta \left( x_t|x_{t-1} \right) \\
    p_\theta \left(y_t|x_{1,\dots,t},y_{T-1} \right) &= g_\theta \left( y_t|x_t \right),
  \end{split}
\end{align}
and assuming that we have calculated $p\left( x_{t-1} | y_{T-1} \right)$ at the previous timestep, we can perform the following calculations:
\begin{align}
  \label{eq:predictionIntegral}
  \begin{split}
    p_\theta \left(x_t | y_{T-1} \right) &= \int p_\theta \left( x_t, x_{t-1} | y_{T-1} \right) dx_{t-1} \\
    &= \int p_\theta \left( x_{t-1} | y_{T-1} \right) f_\theta \left( x_t | x_{t-1} \right) dx_{t-1}
  \end{split} \\
  p_\theta \left(x_t | y_T \right) &= \frac{ p_\theta \left( x_t | y_{T-1} \right) g_\theta \left( y_t | x_t \right) }{ p_\theta \left(y_t | y_{T-1}\right) }. \label{eq:updateBayes}
\end{align}

Derivation of~\eqref{eq:predictionIntegral} and~\eqref{eq:updateBayes} is straightforward; see e.g. \cite{doucet2011tutorial} for more detail. Notice that~\eqref{eq:predictionIntegral} and~\eqref{eq:updateBayes} are the probabilistic formulations of the filtering \emph{prediction} and \emph{update} steps, respectively. Computing the integral in~\eqref{eq:predictionIntegral}, in probabilistic terms, is the act of marginalizing out the variable $X_{t-1}$ from the joint PDF $p_\theta \left(x_t, x_{t-1} | y_{T-1} \right)$. This marginalization is often presented as a ``model update'' where $p_\theta \left(x_t | y_{T-1} \right)$ is found explicitly through equations derived from a model of the PDF $f_\theta(\cdot)$.

Also notice that the update step in~\eqref{eq:updateBayes} is a statement of Bayes' Rule, with $p_\theta \left(x_t | y_t \right)$ being the posterior PDF, $p_\theta \left( x_t | y_{T-1} \right)$ the prior PDF, and $g_\theta \left( y_t | x_t \right)$, the observation function in~\eqref{eq:xAndYAsRVs}, the likelihood (hence the use of alternate names for recursive filtering such as \emph{Bayesian filtering}~\cite{doucet2011tutorial}). The marginal likelihood $p_\theta \left(y_t | y_{T-1}\right)$ plays the role of a normalizing constant.

\subsection{Past work on filtering on macroscopic flow models}
\label{sec:filteringCTM}

We now consider the problem of traffic state estimation on using recursive filtering.
In this paper, we focus on freeways; a large and separate body of literature exists on state estimation for arterial roads.

Most filtering schemes in the literature, as well as the presented in this paper (which relies on the CTM), use Eulerian flow models. As mentioned in Section \ref{sec:macroscopicModels}, Eulerian models come in first- and higher-order varieties. Higher-order models add additional PDEs to~\eqref{eq:LWR} or dynamic equations to the discretization. Various authors have proposed filtering schemes based on both first- (see for example the work of Sun et al.,~\cite{sun2004mixturekalmanfilter} and Work et al.~\cite{work2009trafficmodel}) and second-order models (such as Wang and Papageorgiou~\cite{wang2005EKF} and Mihaylova et al.~\cite{mihaylova2007PF}). In particular, the second-order models of~\cite{wang2005EKF,mihaylova2007PF} add a dynamic equation for the link velocity. While the particular algorithm proposed in Section~\ref{sec:rbpf} uses a first-order model, we remain agnostic on the question of deciding between first- or higher-order models for filtering; the following discussion is intended to be universal for filtering with Eulerian models of any order.


It will provide clarity to the discussion if we separate out the different state subsets in the state vector $x$ individually; for our discussion we will consider the state vector as potentially composed of the density, $x^\rho$, and the velocity, $x^v$ of the links. The vectors of observations of link density and velocity are similarly denoted $y^\rho$ and $y^v$, respectively.

Examination of~\eqref{eq:predictionIntegral} and~\eqref{eq:updateBayes} shows that within the prediction and filtering framework, the only term in which the observations $y$ appear is the likelihood, $g_\theta(y|x)$. Naturally, assimilation of observations, whether density or velocity, requires a model for the likelihood through specification of the observation equation. In particular, the likelihood $g_\theta(y|x) = g_\theta(y^\rho, y^v | x^\rho, x^v)$, which is the joint likelihood of the entire observation vector $y$, must be posed by the practitioner, and its proper form, particularly in representing the relationship between density and velocity observations, is not obvious. Previous authors have used various methods that exploit the structure of their particular prediction framework to reduce the complexity of the likelihood when dealing with multi-state assimilation.

The filtering schemes of Wang and Papageorgiou~\cite{wang2005EKF} and Mihaylova et al.~\cite{mihaylova2007PF} use second-order Eulerian models (note that although the model of Mihaylova et al. used flow and velocity as the state variables, this representation is essentially equivalent to a density-velocity representation through~\eqref{eq:velocity}). With a second-order model, the link density and velocity vectors are separately predicted with different explicit functions of the current state:
\begin{align}
\begin{split}
  x^\rho_t &= \mathcal{F}_{\theta,\rho}(x^\rho_{t-1}, x^v_{t-1}) \\
  x^v_t &= \mathcal{F}_{\theta,v}(x^\rho_{t-1}, x^v_{t-1}).
  \end{split}
  \label{eq:secondOrderTransitions}
\end{align}

Under this construction, the following conditional independence assumption has been made explicitly by~\cite{wang2005EKF,mihaylova2007PF}:
\begin{assumption}[Second-order traffic model assumption]
  The density and velocity states of the network at time $t$ are conditionally independent given the state at time $t-1$. Equivalently, $p_\theta(x_t^\rho, x_t^v | x_{t-1}) = p_\theta(x_t^\rho | x_{t-1}) p_\theta( x_t^v | x_{t-1} )$ \label{assumption1}
\end{assumption}

In our view, Assumption~\ref{assumption1} is not a good assumption. In particular, it seems to conflict with a construction of a second-order model. Without going into too much detail, adding additional PDEs to~\eqref{eq:LWR} implies a belief that vehicle traffic dynamics are too complex to be modeled with~\eqref{eq:LWR} alone, i.e. with local interactions between the state variables $\rho$ and $v$. Discretization for numerical forward integration as in~\eqref{eq:secondOrderTransitions} would then lose these additional interactions, except for simulation timescales on the order of the inter-state interactions.

The schemes of Wang and Papageorgiou~\cite{wang2005EKF} and Mihaylova et al.~\cite{mihaylova2007PF} make another, assumption that is unstated in both~\cite{wang2005EKF} and~\cite{mihaylova2007PF} but is reasonable: that density measurements $y^\rho$ are independent of the velocity state $x^v$ (and vice-versa). One may then factor the likelihood in a straightforward manner:
\begin{align}
    g( y_t | x_t ) &= g(y^\rho_t, y^v_t | x^\rho_t, x^v_t) = p(y^\rho_t | x^\rho_t) p(y^v_t | x^v_t). \label{eq:secondOrderLikelihoodFactoring}
\end{align}
Thus, the ungainly likelihood $g( y_t | x_t )$ need not be specified, and instead only individual observation equations for density and velocity need be specified, the likelihood of the entire observation vector being their product. By beginning with a second-order model that contains both $x^\rho$ and $x^v$ among its states, this likelihood factorization is natural and elementary.

Despite the prevalence of first-order models for macroscopic simulation, an equivalent operation for first-order models, and thus a ``clean'' method for model-based filtering of data from other data domains, including Lagrangian data, is not obvious. If, for example, a first-order model had as its only state a link's density, then the factorization of~\eqref{eq:secondOrderLikelihoodFactoring} is not implementable, as $x^v$ and $p(y^v_t | x^v_t)$ do not exist. As a result, Lagrangian data do not naturally fit into a first-order Eulerian flow model.

We will highlight two works in filtering Lagrangian data in an Eulerian flow model. The first, due to Lovisari et al. \cite{lovisari_data_2015}.
Lovisari et al. choose to stay entirely in the density regime. While we mentioned methods such as the ``g-factor'' algorithm \cite{jia2001gfactor} to calculate densities solely from single-loop Eulerian flow data, Lovisari et al. use probe velocity measurements as substitutes for those from double-loop detectors. If the probe velocities are accurate, density calculations at single-loop detectors can be comparable to those at double-loop detectors. Unfortunately, if single-loop-detector coverage is sparse, usefulness would drastically decrease.

\begin{figure}
	\centering
	\includegraphics[width=.75\columnwidth]{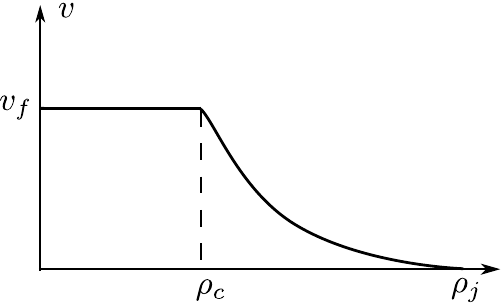}
	\caption{Velocity as a function of density with the Daganzo fundamental diagram (c.f.~\eqref{eq:velocity}). The function is constant for $\rho < \rho_c$ and has a hyperbolic shape for $\rho > \rho_c$. The function is not injective in the region $\rho < \rho_c$.}
	\label{fig:FDvelocity}
\end{figure}

In another work, Work et al.~\cite{work2009trafficmodel}, brought Lagrangian data into Eulerian coordinates in a novel manner. The ``velocity Cell Transmission Model'' (v-CTM) proposed by Work et al. is a first-order model in which the state vector consists only of the velocity along links. That is, rather than typical first-order models that state the sending and receiving functions as functions of link density, the model of Work et al. posed them as functions of velocity. Density and velocity measurements are then fused by transforming density measurements to virtual measurements of the equivalent velocity specified by \eqref{eq:velocity}. While this approach was successful in fusing measurements of different domains (and velocity measurements that are nonfixed), it has its own drawbacks. First, it requires the selection of a fundamental diagram with a bijective relationship between density and velocity~\cite{work2009trafficmodel}, which proscribes many popular fundamental diagrams such as the classic model of Daganzo~\cite{daganzo1994ctm}, whose model of traffic having a constant freeflow velocity is intuitively appealing. In addition, by converting density measurements to velocity measurements, some amount of observed information is lost (specifically, note that~\eqref{eq:velocity} will equate to the freeflow velocity for any value of $\rho < \rho_c$) or contaminated (due to possible modeling errors in the fundamental diagram) (Fig.~\ref{fig:FDvelocity}).

As a consequence, while the method of Work et al. allows estimation of the velocity state trajectory, the density state is not recoverable in general when using the v-CTM. In this paper, we avoid this problem by estimating density directly with our density and velocity measurements.

Thus far we have focused entirely on filtering with Eulerian flow models, but the (relatively sparse) literature on filtering with Lagrangian flow models is worth mentioning. Yuan et al. \cite{yuan_real-time_2012} describe one such algorithm. Their model fuses Eulerian (loop detector measurements) and Lagrangian (high-precision individual vehicle traces) to estimate in the Lagrangian coordinates. A state estimator in the Lagrangian coordinates seeks to reconstruct vehicle ordering and location over time, which may be difficult with contemporary Lagrangian data sources such as low-frequency and noisy GPS. However, with higher-precision Lagrangian data increasing in market penetration, this may soon be feasible.

\subsection{Ensemble methods for macroscopic flow models}
\label{sec:ensembleCTM}

Due to their complex structure, stochastic implementations of macroscopic traffic models create transition kernels $f_\theta(\cdot)$ for which evaluation of the integral in~\eqref{eq:predictionIntegral} is analytically difficult or impossible~\cite{mihaylova2007PF}. The distribution of $X_{t}|X_{t-1}$ is thus not able to be expressed in closed form unless restrictive assumptions are made. These difficulties trace back to the fact that macroscopic traffic models are discretizations of the LWR PDE, leading to a system with states that are tightly coupled with commonly-occuring nonlinear behavior such as congestion and shockwaves. These nonlinearities mean that individual links may variously affect their upstream link, downstream link, or both during the next model update, depending on their and their neighbors' current state. For macroscopic traffic models of moderate or large numbers of links, these cascading nonlinearities can cause large multimodality in distributions~\cite{blandin2012sequential}, which are difficult to approximate in a parametric manner. Indeed, in~\cite{blandin2012sequential}, Blandin et al. showed that a natural approximation of $f_\theta(\cdot)$, the linearized estimate as used in the Extended Kalman Filter, produces estimates that diverge quickly from the true traffic state.

Further, macroscopic traffic models of traffic networks are by construction formed of multiple PDE discretizations that interact with each other through junctions, increasing the occurrence of nonlinearities. Stochastic traffic models for which evaluations of~\eqref{eq:predictionIntegral} is a closed-form operation must make restrictive assumptions to control these nonlinearities, namely assuming high levels of instrumentation stochastically simulating only a freeway, leaving onramps, offramps, and arterial roads as known and nonrandom (see for example \cite{sun2004mixturekalmanfilter} or \cite{sumalee2011stochastic}). While this is effective for simulating well-instrumented freeways, the use of GPS data for traffic estimation is most highly desired in locations with relatively low detection infrastructure, making simplified models undesirable.

For these reasons, \emph{ensemble} or \emph{sequential Monte Carlo} methods have been the major focus of research in traffic state estimation. At a high level, these method also approximate $f_\theta(\cdot)$, but rather than by approximating it with a computationally tractable model, they approximate $f_\theta(\cdot)$ with classical Monte Carlo techniques - sampling repeatedly from $f_\theta \left( x_t|x_{t-1} \right)$ and obtaining the Monte Carlo approximation to~\eqref{eq:predictionIntegral},
\begin{align}
\label{eq:monteCarloPredict}
  \begin{split}
  p_\theta \left(x_t | y_{T-1} \right) &= \int p_\theta \left( x_{t-1} | y_{T-1} \right) f_\theta \left( x_t | x_{t-1} \right) dx_{t-1} \\
  &\approx \sum_{p=1}^P p_\theta \left(x_{p,t-1} | y_{T-1} \right) \delta_{f_\theta} \left(x_{p,t}|x_{p,t-1} \right) \\
  &= \hat{p}_\theta \left(x_t | y_{T-1} \right),
  \end{split}
\end{align}
where $P$ is some integer denoting the total number of samples drawn from $f( \cdot )$, $p \in \{1,\dots,P\}$ indexes individual samples (or atoms of the probability distribution), and $\delta_{f_\theta} \left(x_{p,t}|x_{p,t-1} \right)$ is the Dirac delta, which places a unit mass on the point $x_{p,t}|x_{p,t-1}$, itself denoting the value of the $p$th sample from $f(\cdot)$. The final equality indicates that the empirical PDF $\hat{p} (x_t|y_{T-1})$ consists of a weighted sum of $P$ point masses, with support on the point $x_{p,t}|x_{p,t-1}$, with individual weights $p_\theta(x_{p,t-1}|y_{T-1})$, where the weights sum to one. A straightforward application of the strong law of large numbers shows that as $P \to \infty$, $\hat{p} \left(x_t | y_{T-1} \right) \to p\left(x_t | y_{T-1} \right)$ almost surely~\cite{doucet2011tutorial}.

As our filtering scheme is recursive, the term $p_\theta \left(x_{p,t-1} | y_{T-1} \right)$ denotes the posterior probability assigned to atom $p$ from the previous timestep, which is approximated by replacing $p_\theta(x_t|y_{T-1})$ in~\eqref{eq:updateBayes} with $\hat{p}_\theta(x_t | y_{T-1})$ from~\eqref{eq:monteCarloPredict}:
\begin{align}
	p_\theta(x_t | y_T) &= \frac{p_\theta(x_{t}|y_{T-1}) g_\theta( y_t | x_{t})}{p_\theta (y_t | y_{T-1})} \notag \\
  &\approx \frac{\hat{p}_\theta(x_{t}|y_{T-1}) g_\theta( y_t | x_{t})}{p_\theta (y_t | y_{T-1})} \notag \displaybreak[0] \\
  \begin{split}
  &= \frac{1}{p_\theta(y_t | y_{T-1})} \Bigg[ \sum_{p=1}^P p_\theta \left( x_{p,t} | y_{T-1} \right) \\ 
  &\hphantom{= \frac{1}{p_\theta(y_t | y_{T-1})} \Bigg[} \times g_\theta \left(y_t | x_{p,t}\right) \delta_{f_\theta} \left(x_{p,t}|x_{p,t-1} \right) \Bigg]
  \end{split} \notag \displaybreak[0] \\
  &= \frac{1}{p_\theta(y_t | y_{T-1})} \sum_{p=1}^P p_\theta \left( x_{p,t} | y_{T} \right) \delta_{f_\theta} \left(x_{p,t}|x_{p,t-1} \right) \notag \\
  &= \hat{p}_\theta \left( x_t | y_T \right), 
\label{eq:monteCarloUpdate}
\end{align}
where $g_\theta(y_T | x_{p,t})$ denotes a pointwise evaluation of the likelihood function for the value of the $p$th sample at time $t$.

The factor $1/ p_\theta(y_t | y_{T-1})$ plays the role of a normalizing constant. In practice it is not calculated explicitly. Instead, after the posterior probability $p_\theta \left( x_{p,t} | y_{T} \right)$ is calculated for each particle, the $P$ probabilities are normalized so that their sum equals one. This normalized version of the posterior probabilities $p_\theta \left( x_{p,t} | y_{T} \right)$, and the prior probabilities $p_\theta \left( x_{p,t} | y_{T-1} \right)$, are usually referred to the weight of particle $p$ in their respective empirical PDFs. They are often abbreviated as $w_{p,t-1}$ and $w_{p,t}$, respectively. The Monte Carlo prediction and update steps in~\eqref{eq:monteCarloPredict} and~\eqref{eq:monteCarloUpdate} then become
\begin{subequations}
  \begin{align}
    \hat{p}_\theta \left(x_t | y_{T-1} \right) &= \sum_{p=1}^P w_{p,t-1} \delta_{f_\theta} (x_{p,t} | x_{p,t-1}) \label{eq:PFpredict} \\
    \hat{p}_\theta \left(x_t | y_T \right) &= \sum_{p=1}^P w_{p,t} \delta_{f_\theta} (x_{p,t} | x_{p,t-1}), \label{eq:PFupdate}
  \end{align}
  where the weights are normalized after a measurement is received so the PDF will sum to one as previously discussed:
  \begin{align}
    w_{p,t} = \frac{w_{p,t-1}g_\theta( y_t | x_{p,t})}{\sum_{p} w_{p,t-1} g_\theta( y_t | x_{p,t}) }.
  \end{align}
\end{subequations}
In practice, some portion of the particles will stray very far from the true state, obtain very low weights, and be useless for state estimation. Therefore, after evaluation of a posterior PDF in~\eqref{eq:PFupdate}, a \emph{resampled} version of $\hat{p} (x_t|y_T)$ may be created from $P$ particles of the original empirical PDF, sampled with replacement. Various resampling schemes have been proposed and studied (see, for example, the discussion in~\cite{doucet2011tutorial}), and deep discussion is beyond the scope of this paper. In this paper we use a simple multinomial resampling scheme, where each particle has a selection probability of $w_{p,t}$.

Finally, note that the likelihood $g_\theta( \cdot )$ is only evaluated for specific values of the conditioned term - namely, the particle-specific value $x_{p,t}$. Let us make the following conditional independence assumption to take advantage of this:
\begin{assumption}
  Given a value of $X_t$, the state of the entire network at time $t$, individual measurements in the vector $Y_t$ are conditionally independent. Equivalently, the measurement noises of individual measurements are independent.
  \label{assumption2}
\end{assumption}

The above assumption allows us to further factor the likelihood:
\begin{equation}
\begin{split}
  g_\theta( y_t | x_{p,t} ) &= \prod_{i=1}^M p_\theta( y_{t,i} | x_{p,t} ) \\
  &= \prod_{i=1}^M p_\theta( y_{t,i} | x_{p,t,L(i)} ),
\end{split}
\end{equation}
where $i \in \{1,\dots,M \}$ indexes individual elements of the $M$-long measurement vector $y$ and $L(i)$ denotes the link where measurement $i$ takes place.

Given an ensemble of particles $\{x_{t,p}|y_{t,p}\}_{p=1}^P$, a variety of point estimates of the system state may be obtained, among them the empirical mean
\begin{align}
  \mathbf{E}\left[ X_t|Y_t \right] \approx \sum_{p=1}^P w_{t,p} \delta \left(x_{p,t}|y_t \right). \label{eq:empiricalMean}
\end{align}

The above construction is implemented to estimate density using only density measurements in Algorithm~\ref{alg:PF}. Algorithm~\ref{alg:PF} is itself not novel.
\newtheorem{alg}{Algorithm}

\begin{alg}[Particle Filter for Traffic Density Estimation]
\leavevmode
\mbox{}

\noindent Inputs: 
\begin{itemize}
  \item PDF of density initial conditions $p_\theta(X^\rho_0)$
  \item Stochastic cell transmission model $f_\theta( x^\rho_{p,t} | x^\rho_{p,t-1})$
  \item Density measurement likelihood function $p_\theta( y_i | x^\rho_{L(i)} )$, where $L(i)$ indicates the link of the $i$th measurement
\end{itemize} \mbox{}

  \begin{enumerate}
  \item Initialization: At time $t = 0$:
  \begin{enumerate}
    \item Sample an ensemble of particles of the density state, $x^\rho_{p,0} \sim p_\theta(X^\rho_0)$, the distribution for the initial condition \\ $\forall \, p \in \{1,\dots,P\}.$
    \item Set the initial weights $w_{p,0} = 1/P \; \\ \forall \, p \in \{1,\dots,P\}.$
  \end{enumerate}
  
  \item Prediction: At time $t > 0$, $\forall$ particle $p$, sample \\ $x^\rho_{p,t} \sim f_\theta( x^\rho_{p,t} | x^\rho_{p,t-1})$ by evaluating a one-step stochastic cell transmission model update \label{steppredDensity}
  \item Data assimilation: $\forall$ particle $p$, at time t
  \begin{enumerate}
    \item For each measurement received at time t, $y_{t,i}$, compute the per-measurement likelihood \\ $g_{p,i} = p( y_{t,i} | x^\rho_{p,t,L(i)})$
    \item Compute the overall particle likelihood, \\ $g_p = \prod_i g_{p,i}$
    \item Compute the (unnormalized) posterior particle weight, $\tilde{w}_{p,t} = w_{p,t-1} g_p$
  \end{enumerate}
  \item Normalization: Normalize the particle weights, \\ $w_{p,t} = \tilde{w}_{p,t}/ \left( \sum_p \tilde{w}_{p,t} \right)$
  \item Resampling: If resampling is desired, resample $P$ particles with replacement from $\{x^\rho_{p,t}\}$ with selection probability of particle $p = w_{p,t}$
  \item If $t = t_{final}$, end, otherwise $t \leftarrow t+1$ and return to step \ref{steppredDensity}
  \end{enumerate}

\label{alg:PF}
\end{alg}

\section{Rao-Blackwellized Particle Filter for Data Fusion}
\label{sec:rbpf}

\subsection{Rao-Blackwellization as an improvement for sequential Monte Carlo}

Our algorithm makes use of a modification to standard particle filtering known as a \emph{Rao-Blackwellized Particle Filter} (RBPF). The name refers to the Rao-Blackwell Theorem, a well-known result from mathematical statistics (see e.g. the discussion in \cite[Ch.~3]{keener2010theoretical}), which states that an estimate based on data may be improved in terms of expected convex loss, and will never be worsened, by conditioning on a sufficient statistic. In the setting of Monte Carlo methods, \emph{Rao-Blackwellization} refers to a method for improving a Monte Carlo sampler over several random variables. Should some subset of the random variables have distributions as explicit functions of another subset, gains in computational cost and accuracy can be made by making use of these explicit distributions, rather than approximating them with Monte Carlo~\cite{casella1996raoblackwell}. Rao-Blackwellization has gained broad adoption in Sequential Monte Carlo methods in particular~\cite{doucet2000rbpf}. If there exists some subset of the state vector, $x^B$ whose distribution is an explicit function of the remaining states, $x^A$ then by analogy $x^A$ is a sufficient statistic for $x^B$. We may then take a shortcut in evaluating our Monte Carlo prediction step in~\eqref{eq:monteCarloPredict}:
\begin{align}
  &p_\theta \left(x_t | y_{T-1} \right) \nonumber \\ 
  &= \int p_\theta \left( x_{t-1} | y_{T-1} \right) f_\theta \left( x_t | x_{t-1} \right) dx_{t-1} \nonumber \\
  &= \int  p_\theta \left( x_{t-1} | y_{T-1} \right) f_{\theta, A}\left( x^A_t | x_{t-1} \right) f_{\theta, B} \left(x^B_t|x^A_t\right) dx_{t-1} \nonumber \\
  &\approx \sum_{p=1}^P p_\theta( x_{p,t} | y_{T-1}) \left( \delta_{f_\theta,A}\left( x^A_{p,t} | x_{p,t-1} \right) \times
  \delta_{f_\theta,B} \left(x^B_{p,t}|x^A_{p,t}\right) \right), \label{eq:raoBlackwellPredict}
\end{align}
where $f_{\theta,A}\left(\cdot\right)$ is a PDF for the state subset $X^A_t | X_{t-1}$ (that is, a truncated version of $f_\theta(\cdot)$ that only has domain in the space of $x^A$), $f_{\theta,B}(\cdot)$ is our closed-form PDF for $x^B$, and the use of the Cartesian product $\times$ is needed due to the individual Dirac deltas residing on disjoint subsets of $x$. In addition, the second line makes use of the fact that, $X^B$ being an explicit function of $X^A$, we have that $X^B_t$ is conditionally independent of $X_{t-1}$ given $X^A_t$.

When implementation is possible, Rao-Blackwellization of a particle filter offers significant advantages. Reducing the size of the state that one must approximate through Monte Carlo brings improvements in computation time and eliminates error in approximating $x^B$ through Monte Carlo approximation. Indeed, Rao-Blackwellization of particle filters were originally proposed for a reduction in the variance of estimated distributions~\cite{doucet2000rbpf}. Reducing computation time also allows for improved estimates by freeing additional computational resources to simulating more particles, allowing for richer predictions of $p\left(x_t | y_{T-1}\right)$. For these reasons, RBPFs have gained recent popularity for improving approximations of computationally difficult problems in large state-space settings, such as the simultaneous localization and mapping problem in robotics~\cite{grisetti2007rbpfSLAM}. We are particularly interested, though, in its immediate application for data fusion.

\subsection{Implementation for traffic data fusion}

As mentioned in Section \ref{sec:ensembleCTM}, first-order traffic models predict one-step model updates only in terms of density, leaving the velocity likelihood term $p_\theta \left( y_t^v | x_t^v \right)$ in~\eqref{eq:secondOrderLikelihoodFactoring} unspecified. The lack of a plug-in likelihood has been a hindrance for data fusion on first-order models, but by making use of an RBPF, we may overcome this.

Recall from \eqref{eq:velocity} that under a first-order traffic model, the average velocity of a link may be computed from the link's density and flow (itself a function of the density). Let us then say that link density is a sufficient statistic in the Rao-Blackwell sense for link velocity. We may then factor the likelihood in an analogous manner to the operation in~\eqref{eq:secondOrderLikelihoodFactoring},
\begin{align}
  g_\theta( y_t | x_t ) &= g_\theta( y_t^\rho, y_t^v | x^\rho_t, \bar{x}_t^v) = p_\theta( y_t^\rho | x_t^\rho ) p_\theta( y_t^v | \bar{x}_t^v, x_t^\rho ) \nonumber \\
  &= p_\theta( y_t^\rho | x_t^\rho ) p_\theta( y_t^v | \bar{x}_t^v ) p_\theta( \bar{x}_t^v | x_t^\rho ), \label{eq:firstOrderLikelihoodFactoring}
\end{align}
where we introduce $\bar{x}^v_t$, a random variable denoting the average velocity of a link, whose PDF $p_\theta( \bar{x}^v_t | x^\rho_t )$, is an explicit function of the link density under a first order model, and $p_\theta( y_t^v | \bar{x}_t^v )$ is a likelihood denoting the distribution of velocity measurements given an average velocity. This last function will then incorporate information such as the distribution of vehicle velocities about a link's nominal velocity, as well as measurement noise of individual probe measurements around their sensors' true velocity.

The variable $\bar{x}^v_t$ is a time-varying quantity that describes the system, but it is not a state of the system - the only system state of the CTM is density. It is instead an intermediary between the true state $x^\rho$ and observations $y^v$. We refer to it as a \emph{pseudostate}.

Note that while we have referred to \eqref{eq:velocity} as the source of the PDF $p_\theta( \bar{x}^v_t | x^\rho_t )$, which follows the original formulation \eqref{eq:velocity} in treating link velocity as a nonrandom function of density, various authors including, notably, Sumalee et al. in~\cite{sumalee2011stochastic}, have proposed flow and velocity as random functions of link density. A practitioner would be free to incorporate this stochasticity in this function.

By applying assumption~\ref{assumption2}, we may further simplify~\eqref{eq:firstOrderLikelihoodFactoring}:
\begin{equation}
\begin{split}
  g_\theta( y_t | x_t ) &= p_\theta( y_t^\rho | x_t^\rho ) p_\theta( y_t^v | \bar{x}_t^v ) p_\theta( \bar{x}_t^v | x_t^\rho) \\
  &= \prod_{i=1}^{M^\rho} p_\theta ( y_{t,i}^\rho | x_{t,L(i)}^\rho ) \\
  & \quad \times \prod_{j=1}^{M^v} p_\theta( y_{t,j}^v | \bar{x}_{t,L(j)}^v ) p_\theta( \bar{x}_{t,L(j)}^v | x_{t,L(j)}^\rho ),
\end{split}
\label{eq:firstOrderPerLinkFactoring}
\end{equation}
where $i \in \{1, \dots, M^\rho\}$ indexes the $M^\rho$-long density measurement vector $y_t^\rho$, $j \in \{1,\dots,M^v\}$ performs the same function for $y_t^v$, and $L(\cdot)$ denotes the link of the associated measurement. Equation \eqref{eq:firstOrderPerLinkFactoring} is the likelihood we use in implementing our RBPF.

An implementation of our RBPF scheme is described in algorithm~\ref{alg:RBPF}. This algorithm is constructed in similar manner to algorithm~\ref{alg:PF}, but with ~\eqref{eq:firstOrderPerLinkFactoring} used for the likelihood $g_\theta( \cdot )$.

\begin{alg}[Rao-Blackwellized Particle Filter for Density Estimation on Pseudo-Second-Order Model]
\label{alg:RBPF}
\leavevmode
\mbox{}

\noindent Inputs: 
\begin{itemize}
  \item PDF of density initial conditions $p_\theta(X^\rho_0)$
  \item Stochastic cell transmission model $f_\theta( x^\rho_{p,t} | x^\rho_{p,t-1})$
  \item Per-link predicted velocity distribution $p_\theta( \bar{x}_{t,L(i)}^v | x_{t,L(i)}^\rho )$
  \item Density and velocity measurement likelihood functions $p_\theta^\rho( y_i | x^\rho_{L(i)} )$ and $p_\theta^v( y_i | x^v_{L(i)} )$, where $L(i)$ indicates the link of the $i$th measurement
\end{itemize} \mbox{}

  \begin{enumerate}
  \item Initialization: At time $t = 0$:
  \begin{enumerate}
    \item Sample an ensemble of particles of the initial density, $x^\rho_{p,0} \sim p_\theta(X^\rho_0)$, the distribution for the initial condition \\ $\forall \, p \in \{1,\dots,P\}.$
    \item Set the initial weights $w_{p,0} = 1/P \; \\ \forall \, p_\theta \in \{1,\dots,P\}.$
  \end{enumerate}
  
  \item Prediction: At time $t > 0$, $\forall$ particle $p$, sample \\ $x^\rho_{p,t} \sim f_\theta( x^\rho_{p,t} | x^\rho_{p,t-1})$ by evaluating a one-step stochastic cell transmission model update \label{steppred}
  \item Data assimilation: $\forall$ particle $p$ at time $t$,
  \begin{enumerate}
    \item For each measurement received at time $t$, $y_{t,i}$
    \begin{enumerate}
      \item If $y_{t,i}$ is a density measurement, compute the per-measurement likelihood \\ $g_{p,i} = p_\theta^\rho( y_{t,i} | x^\rho_{p,t,L(i)})$
      \item If $y_{t,i}$ is a velocity measurement, 
      \begin{enumerate}
	\item Compute the predicted link velocity distribution $p_\theta( \bar{x}_{t,L(i)}^v | x_{t,L(i)}^\rho )$
	\item Compute the per-measurement likelihood $g_{p,i} = p_\theta^v( y_{t,i} | \bar{x}^v_{L(i)})$
      \end{enumerate}
    \end{enumerate}
    \item Compute the overall particle likelihood, \\ $g_p = \prod_i g_{p,i}$
    \item Compute the (unnormalized) posterior particle weight, $\tilde{w}_{p,t} = w_{p,t-1} g_p$
  \end{enumerate}
  \item Normalization: Normalize the particle weights, \\ $w_{p,t} = \tilde{w}_{p,t}/ \left( \sum_p \tilde{w}_{p,t} \right) $
  \item Resampling: If resampling is desired, resample $P$ particles with replacement from $\{x^\rho_{p,t}\}$ with selection probability of particle $p = w_{p,t}$
  \item If $t = T$, end, otherwise $t \leftarrow t+1$ and return to step \ref{steppred}
  \end{enumerate}

\end{alg}

\subsection{Discussion}

We refer to this traffic model construction as a \emph{pseudo-second-order} traffic model, due to the middle ground it occupies between traditional first and second-order models. In the traffic data fusion literature, second-order models are proposed due to a claimed ability to estimate velocity as well as density (see e.g.~\cite{mihaylova2007PF}). What is meant is that in a second-order model, the traffic velocity undergoes stochastic updates independent of the traffic density (recall ~\eqref{eq:secondOrderTransitions}). Our model also treats velocity as a random variable and seeks to estimate it, but no Monte Carlo steps are performed in the velocity domain. Instead, velocity state estimation is done in closed form (the Rao-Blackwellization computation of~\eqref{eq:firstOrderLikelihoodFactoring}-\eqref{eq:firstOrderPerLinkFactoring}). The estimation of velocity with a PF does not in general require that the estimation be done with Monte Carlo. Indeed, this recalls the original justification for Rao-Blackwellization of sampling schemes~\cite{casella1996raoblackwell}, in that unnecessary Monte Carlo should be avoided when possible.

Of course, the first-order assumption underlying our use of $x^\rho$ as a ``sufficient'' estimator for $x^v$ in the pseudo-second-order model may reasonably be called into question. There is much debate among traffic theorists as to which PDEs truly govern traffic flow. Our position is that, while it may be the case that the CTM is a simplification that cannot capture some traffic phenomena, its simplicity is a boon in corridor-scale filtering applications. The problem considered in this paper, fusing data from many loop detectors and probe points presupposes application at the scale of a corridor or larger, where more complex models may be unwieldy. As an example, the CTM parameters can be quickly estimated from loop detector data that is obtained in the form of flow-density pairs, but parameters governing additional complex behavior would require additional data sources or hand-tuning, which may become infeasible for corridor-scale applications. Note that the most successful filtering algorithm for traffic data fusion at a corridor scale has been the first-order v-CTM model~\cite{work2009trafficmodel, patire2015howmuchdata}.

Finally, while the development in this section has dealt with a Rao-Blackwellization of a particle filter in particular, note that the derivation only used the PF formalism in specifying the form of the approximation PDFs $\hat{p}_\theta(x_t | y_{T-1})$ and $\hat{p}_\theta(x_t | y_T)$ as weighted sums of Dirac deltas. The Rao-Blackwellization steps can easily be applied to data assimilation with other traffic data filtering schemes by repeating the factorization of the likelihood.

\section{Experimental Results}
\label{sec:experiments}

We describe two numerical experiments to demonstrate the ability of the RBPF in assimilating velocity measurements to improve density estimates. The first of the experiments uses simulated data, where a ``ground truth'' state trajectory was created through stochastic simulation, and PFs were used to recover the full trajectory given fixed-location noisy density measurements $\{y^\rho\}$, simulated moving velocity measurements $\{y^v\}$, or both. The second experiment uses real density and velocity measurements collected from the site, and attempts to predict the density observed by a held-out subset of the loop detectors.

\begin{figure}
  \centering
  \includegraphics[width=20pc]{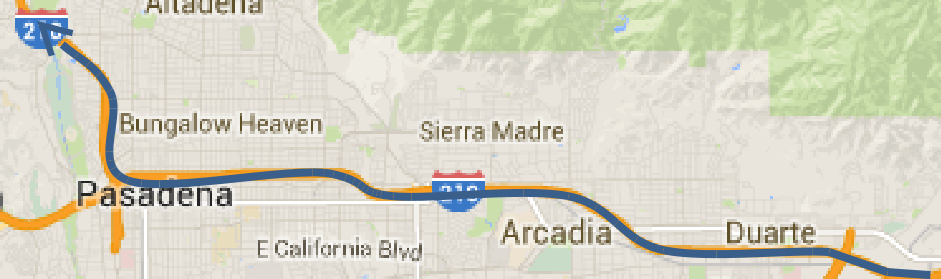}
  \caption{Test site of interest: a 19-mile stretch of I-210W near Los Angeles, CA.}
  \label{fig:210W}
\end{figure}

\begin{figure}
  \centering
  \includegraphics[width=.5\textwidth, trim=0 0 2pc 1.5pc,clip]{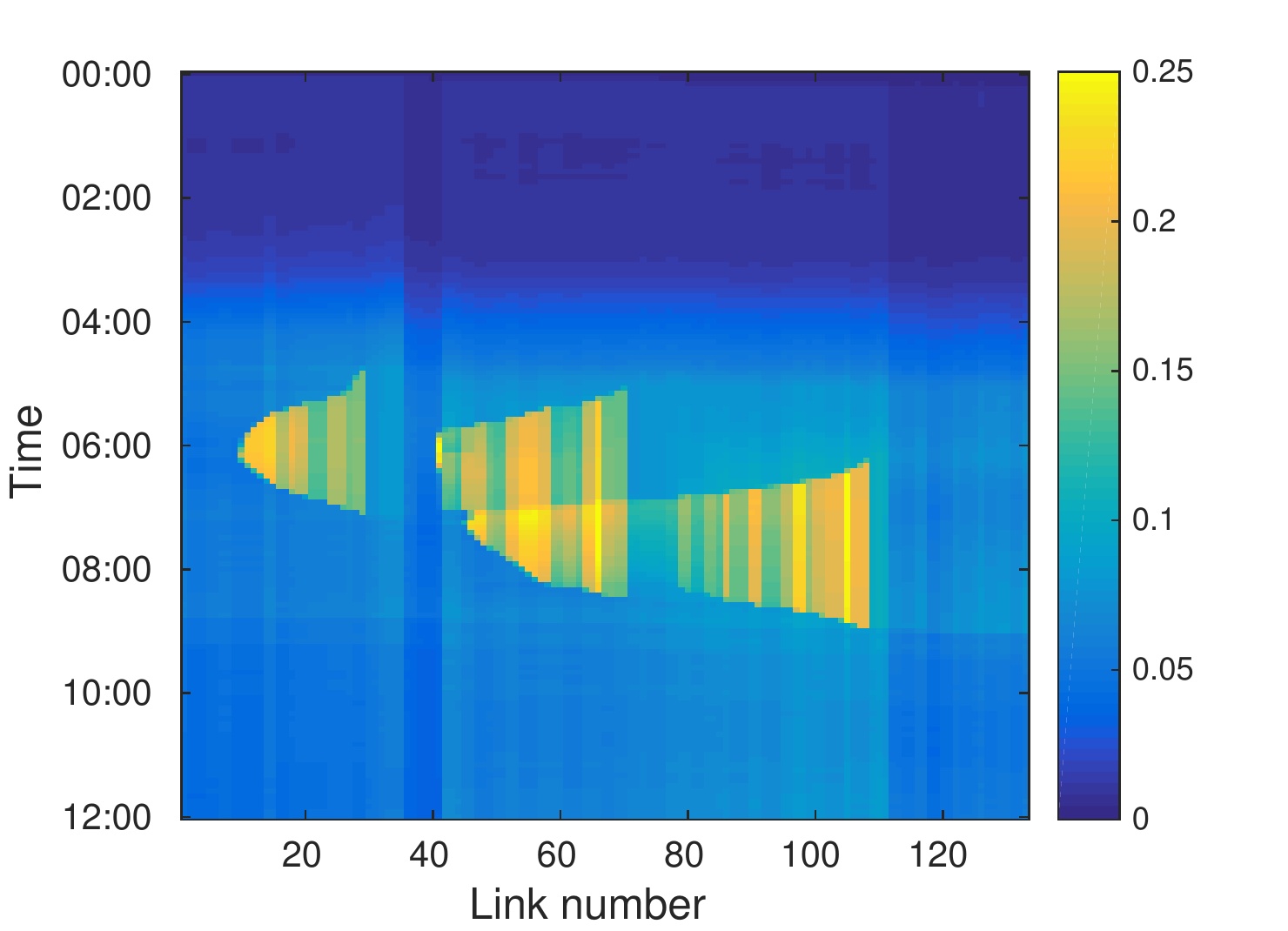}
  \caption{Traffic congestion patterns of the model as fit to October 13, 2014.}
  \label{fig:basicDensity}
\end{figure}

Both experiments use a CTM scheme as the prediction framework, with the model based on a section of Westbound I-210 (Fig. \ref{fig:210W}). The stretch of freeway of interest has a length of approximately 19 miles, and was divided into 127 model links with length averaging roughly 200 m. In addition, the section of interest has 23 onramps and 21 offramps. The stretch of road has 42 PeMS loop detectors~\cite{pems_web} along the freeway mainline. The model was trained against loop data collected during the morning of October 13, 2014. On this date, 8 of the loop detectors were determined to be malfunctioning on Oct 13 through identification as such by the PeMS software or manual checking~\cite{gomes2015i210Wdetection}. The fundamental diagram used in the model was the triangular fundamental diagram due to Daganzo discused in Section~\ref{sec:macroscopicModels}. The fundamental diagram parameters were trained according to the procedures described in~\cite{munoz2004ctmcalibration} and~\cite{dervisoglu2009FDcalibration}, with undetected ramp data imputed according to the method of~\cite{muralidharan2009imputation}. Driver behavior at offramp junctions was modeled with turning ratios, calibrated according to the procedure of~\cite{wright2015splits} from the previous month of data. This direction is the peak direction in a typical morning, so the extent of congestion is relatively large (Fig.~\ref{fig:basicDensity}).

For all experiments, the fundamental diagrams (and thus the PDF $p_\theta(\bar{x}^v_t | x_t^\rho)$) were taken to be noiseless (i.e. a nonrandom function equal to~\eqref{eq:velocity}) for simplicity. A commonly-used likelihood in the literature for assimilating density measurements is a Gaussian density~\cite{wang2005EKF}, we select $g_\theta \left( y^\rho_i|x^\rho_{L(i)} \right) = \mathcal{N}( x^\rho_{L(i)}, \hat{\sigma^2}^\rho_{L(i)} )$ with $\hat{\sigma^2}^\rho_{L(i)}$ being the sample variance of the density of link $L(i)$ across all particles. We also select the likelihood for assimilating velocity measurements as Gaussian, $p_\theta \left( y^v_i|\bar{x}^v_L(i) \right) = \mathcal{N}( x^v_L(i), \hat{\sigma^2}^v_{L(i)} )$ with $\hat{\sigma^2}^v_{L(i)}$ defined similarly.

\subsection{Estimating simulated traffic densities}

\begin{figure*}
  \centering
    \subfloat[]{
  	  \includegraphics[width=20pc,trim=0 0 2pc 1.5pc,clip]{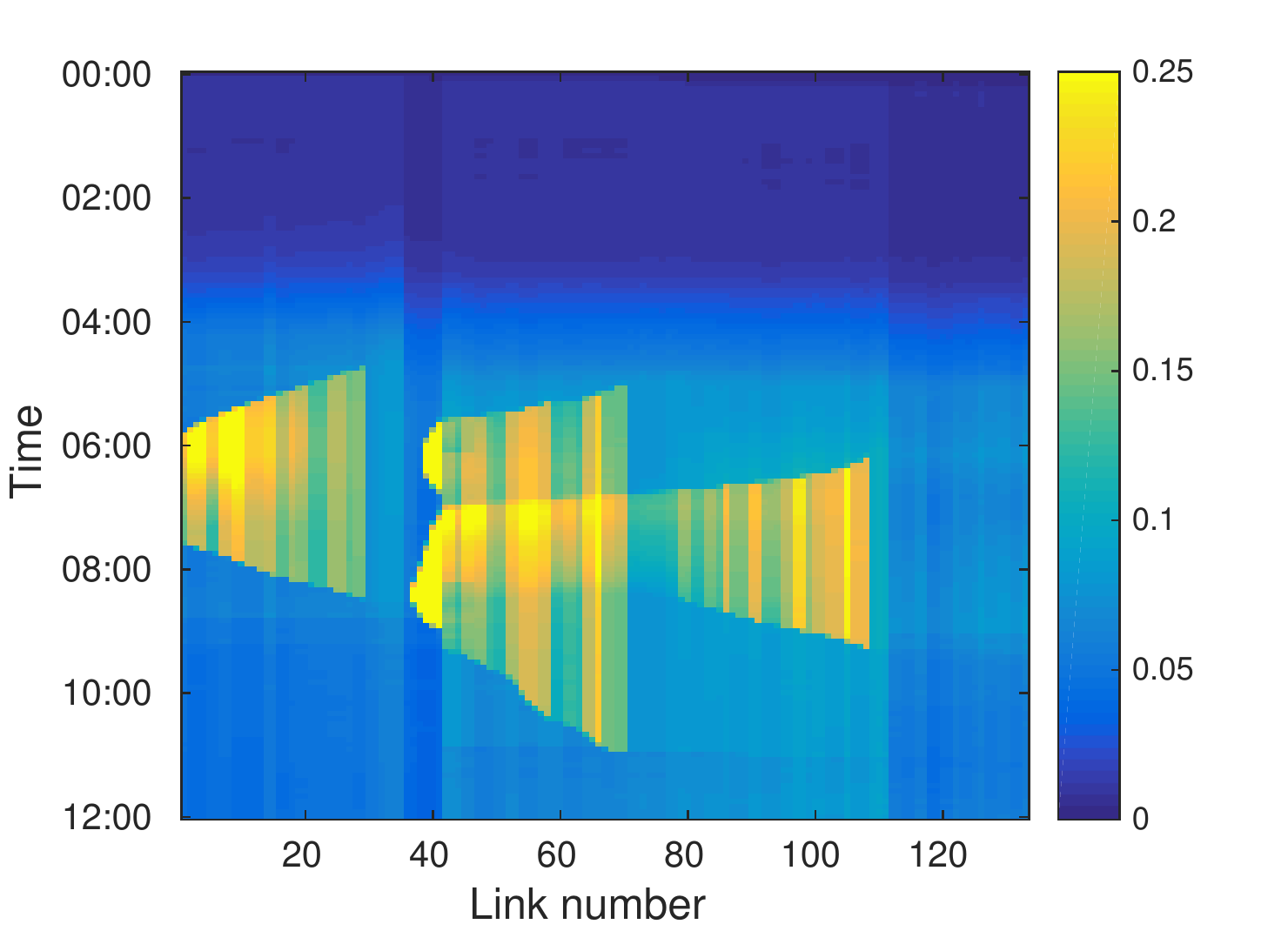}
      \label{fig:simulatedTrue}}
  \subfloat[]{
    \includegraphics[width=20pc,trim=0 0 2pc 1.5pc,clip]{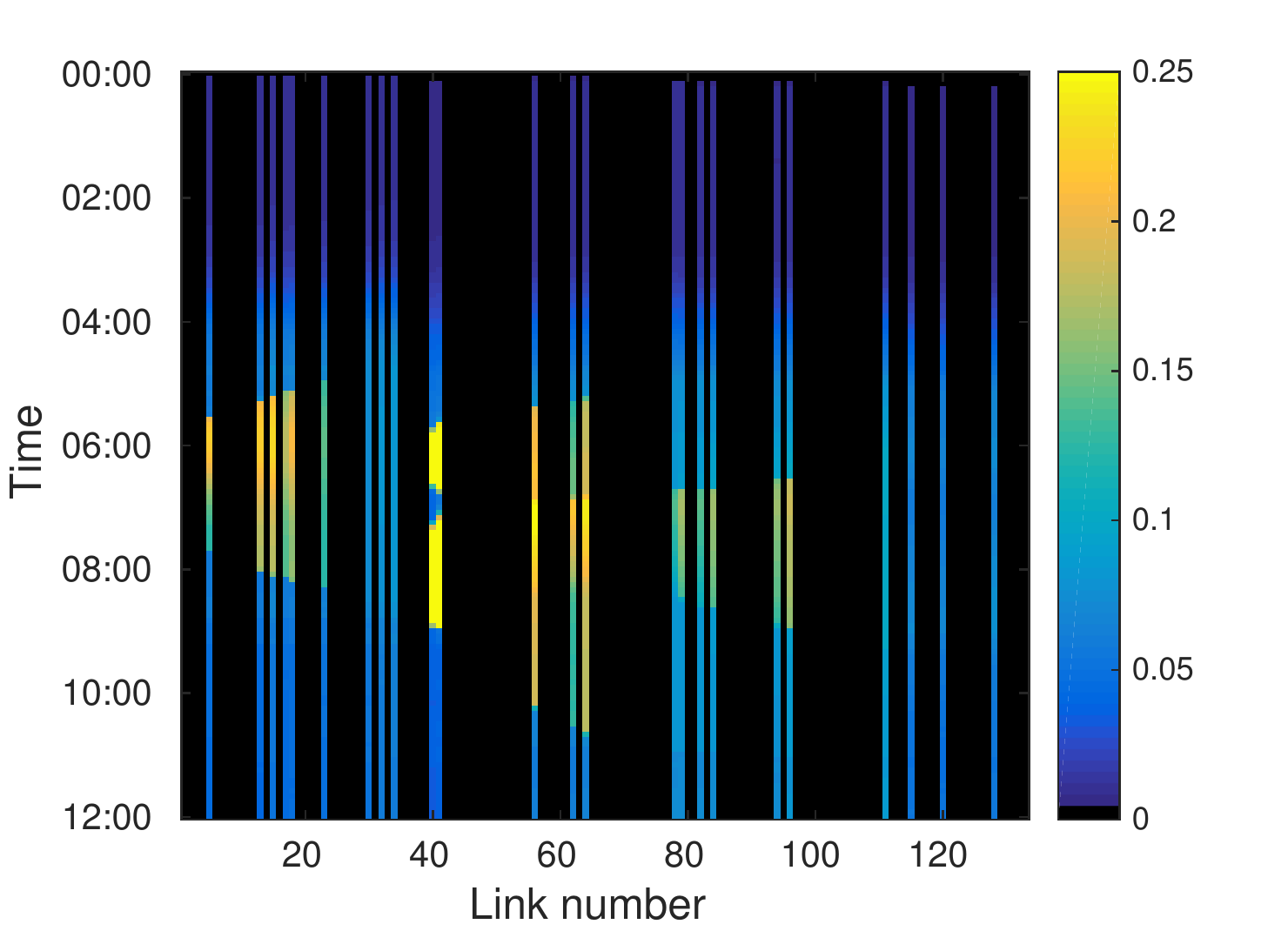}
    \label{fig:simulatedDensity}} \\
  \subfloat[]{
    \includegraphics[width=20pc,trim=0 0 2pc 1.5pc,clip]{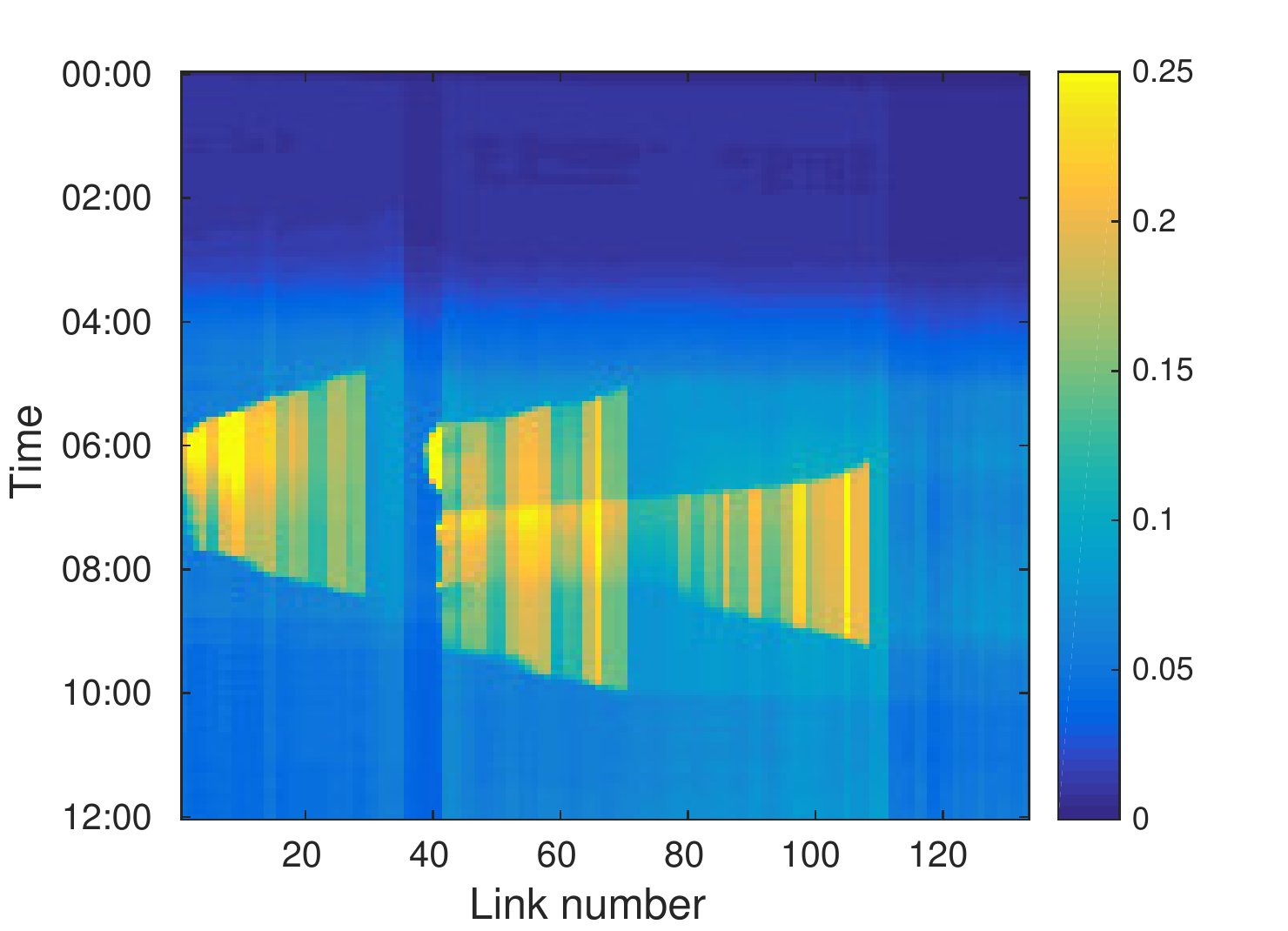}
    \label{fig:simulatedEstimatedDensity}}
  \subfloat[]{
    \includegraphics[width=20pc,trim=0 0 2pc 1.5pc,clip]{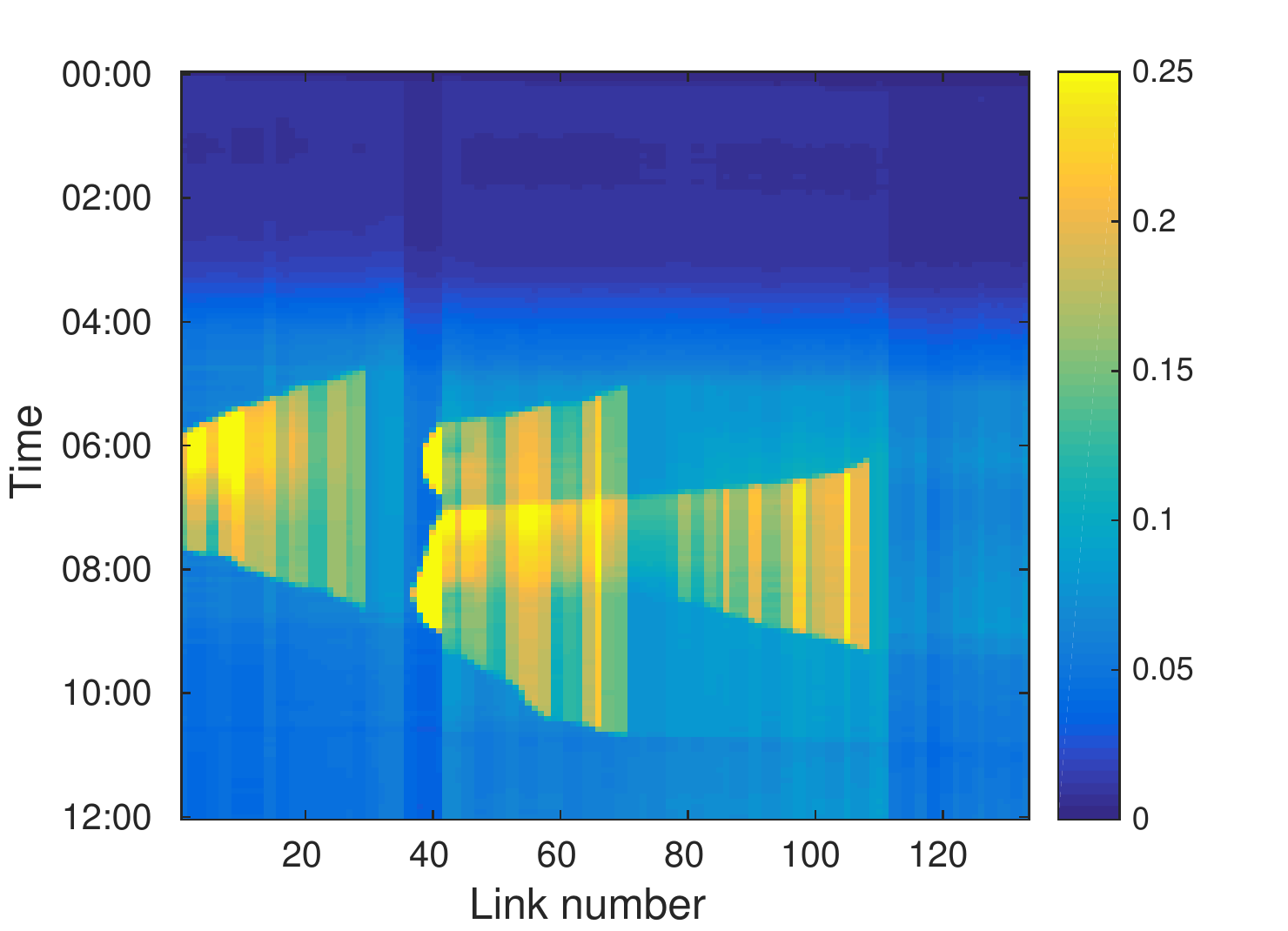}
    \label{fig:simulatedEstimatedProbes}} \\
  \subfloat[]{
    \includegraphics[width=20pc,trim=0 0 2pc 1.5pc,clip]{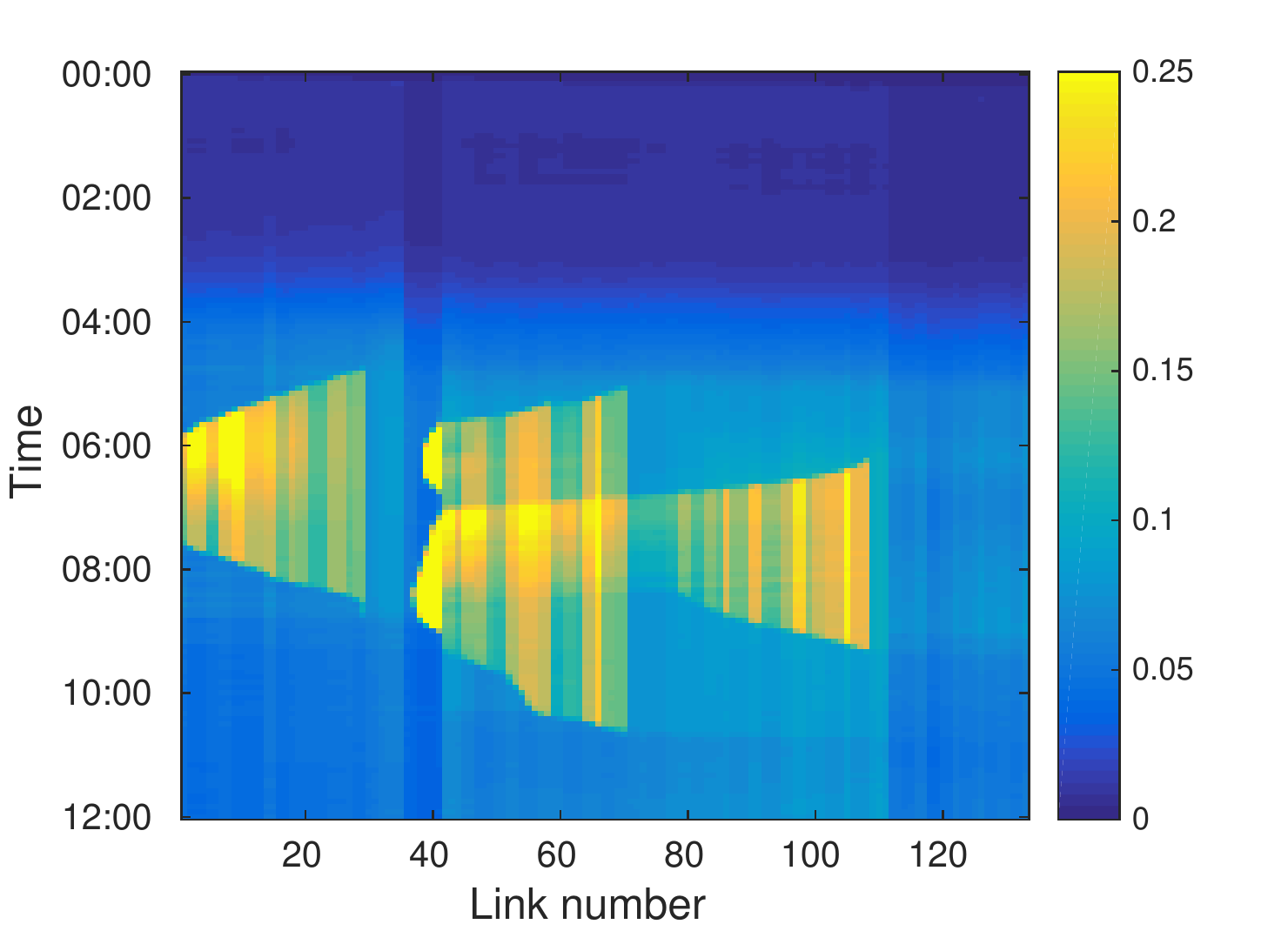}
    \label{fig:simulatedEstimatedBoth}}
  \caption{Density contour maps from a simulated experiment with loop and probe data (veh/m). Black = no data. (a) ``Ground truth'' simulation. (b) Simulated density measurements. (c) Estimated with density measurements. (d) Estimated with velocity measurements of 3\% penetration rate. (e) Estimated with fused data.}
\end{figure*}

This simulation evaluated the performance of the RBPF on reconstructing a known true state. In this experiment, a single simulation of an afternoon period with the stochastic CTM model trained as described above was recorded and taken as the ``actual'' value of $\{x_t\}, t \in 1,\dots,t_{final}$. This ``actual'' value is shown in Fig.~\ref{fig:simulatedTrue}, and featured significantly more congestion than the model baseline. We attempted to estimate the full density state across all timesteps using simulated noisy measurements. The point estimates of $\{x_t\}$ described below are the empirical means~\eqref{eq:empiricalMean} produced by the filter.

The true density state is shown in Fig.~\ref{fig:simulatedTrue}. One estimation run used a particle filter as described above, with access to density measurements of freeway links where detectors are located in the real world (Fig.~\ref{fig:simulatedDensity}). These measurements were contaminated with additive noise sampled from a Gaussian distribution with a standard deviation equal to 10\% of the measured value. The second estimation run used a RBPF and had access to these same density measurements as well as simulated velocity measurements. The velocity measurements were sampled randomly to simulate various penetration rates: for penetration rate $PR$, velocity measurements of number equal to $floor(PR \cdot 100)$ were reported to the filter every five minutes. Each link had a probability proportional to its occupancy of reporting a noisy velocity. This random selection was done with replacement, so a link may report multiple measurements. These five-minute bins were used to mimic the typical procedures of data fusion methods, where probe measurements are retained and filtered into the model at the same time that loop data is next received; for PeMS, these data are reported in five-minute intervals. These velocity measurements were also contaminated with additive noise from a Gaussian distribution with a standard deviation of 10\% of the measured value.

The state estimate generated by the particle filter with access only to these density measurements is shown in Fig. \ref{fig:simulatedEstimatedDensity}, the estimate produced using only the velocity measurements appears in Fig. \ref{fig:simulatedEstimatedProbes}, and the estimate generated by the RBPF with access to both the density and velocity measurements is shown in Fig. \ref{fig:simulatedEstimatedBoth}. We quantified the estimation accuracy by comparing the mean absolute percentage error (MAPE), i.e. the average of the quantities ${\left|\hat{x}_{\ell,t}^\rho - x_{\ell,t}^\rho\right|}/{x_{\ell,t}^\rho}$ for all links $\ell$ and times $t$, of the two runs. The results are summarized in Table \ref{tab:simulatedMAPE}. Examination of the figures and table show that qualitatively, all estimates compare well to the true state, but data fusion via the RBPF quantitatively outperforms the other two estimates in predicting density.

\begin{table*}
  \caption{Density MAPE metrics from simulation. Cong. = Congested, FF = Freeflow.}
  \label{tab:simulatedMAPE}
  \centering
  \begin{tabular}{ r | l l l | l l l }
    & \multicolumn{6}{c}{MAPE} \\
    \cline{2-7}
    & \multicolumn{3}{c|}{Without Simulated Loops} & \multicolumn{3}{c}{With Simulated Loops} \\
    \cline{2-7}
    Simulated Penetration Rate & Cong. Links & FF Links & Overall & Cong. Links & FF Links & Overall \\
    \hline
    None & 28.57\% & 6.54\% & 11.05\% & 8.60\% & 4.00\% & 4.94\% \\
    1\% & 3.86 & 5.64 & 5.28 & 3.65 & 3.72 & 3.71 \\
    2\% & 2.89 & 5.66 & 5.09 & 3.39 & 3.69 & 3.63 \\
    3\% & 2.44 & 5.88 & 4.93 & 2.52 & 3.72 & 3.47 \\
  \end{tabular}
\end{table*}

\subsection{Real data on a corridor scale}
\label{sec:realData}

We now present the results of our real-data experiment. As mentioned above, our model was trained against loop data from October 13, 2014. The dataset used for the experiment was recorded on October 22, 2014. The loop data used were obtained from the California PeMS database~\cite{pems_web}. The probe data were obtained from a major mapping data provider.

\begin{figure*}
  \centering
  \subfloat[]{
    \includegraphics[width=20pc,trim=0 0 2pc 1.5pc,clip]{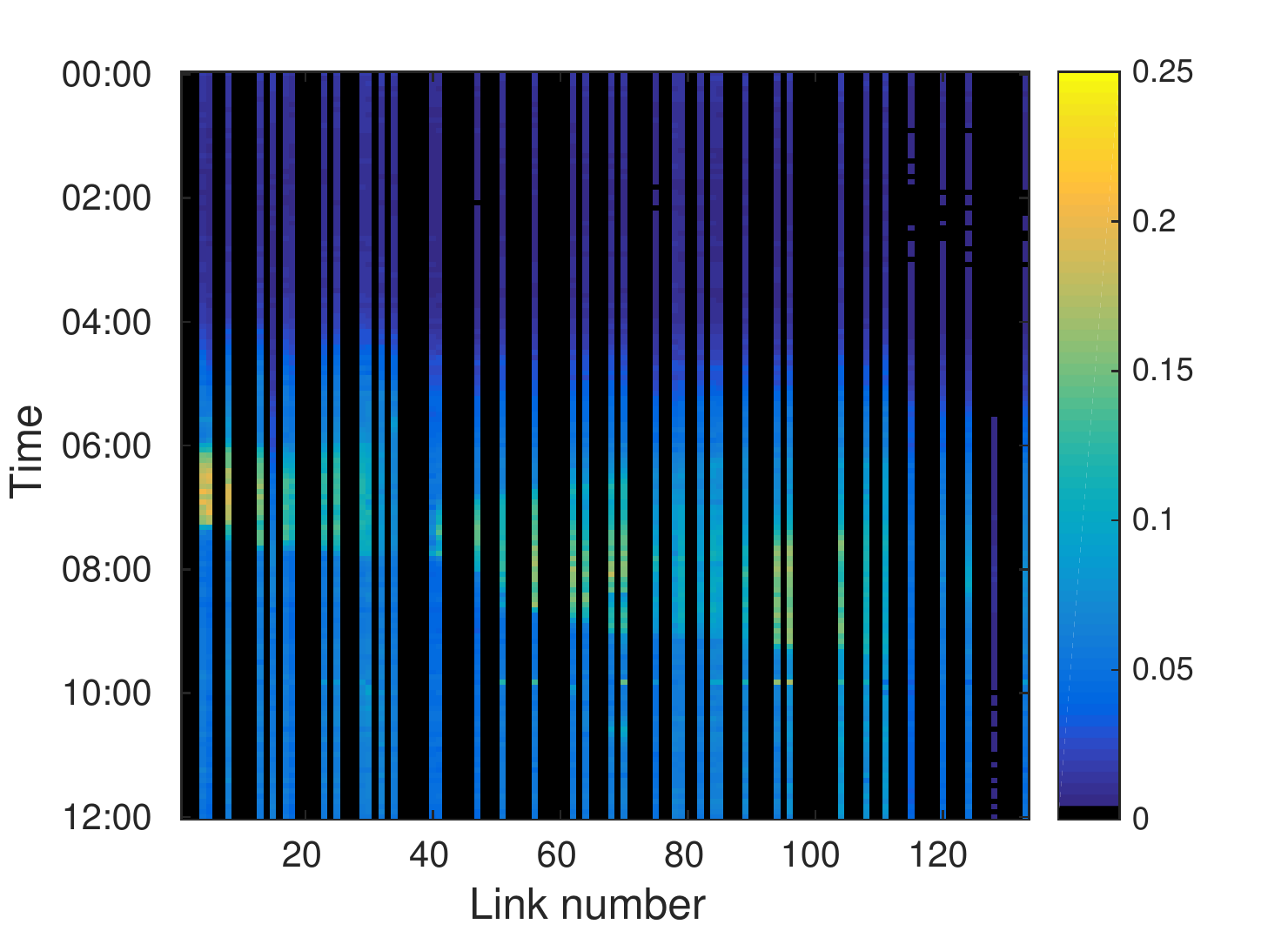}
    \label{fig:Oct13Loops}} 
  \subfloat[]{
    \includegraphics[width=20pc,trim=0 0 2pc 1.5pc,clip]{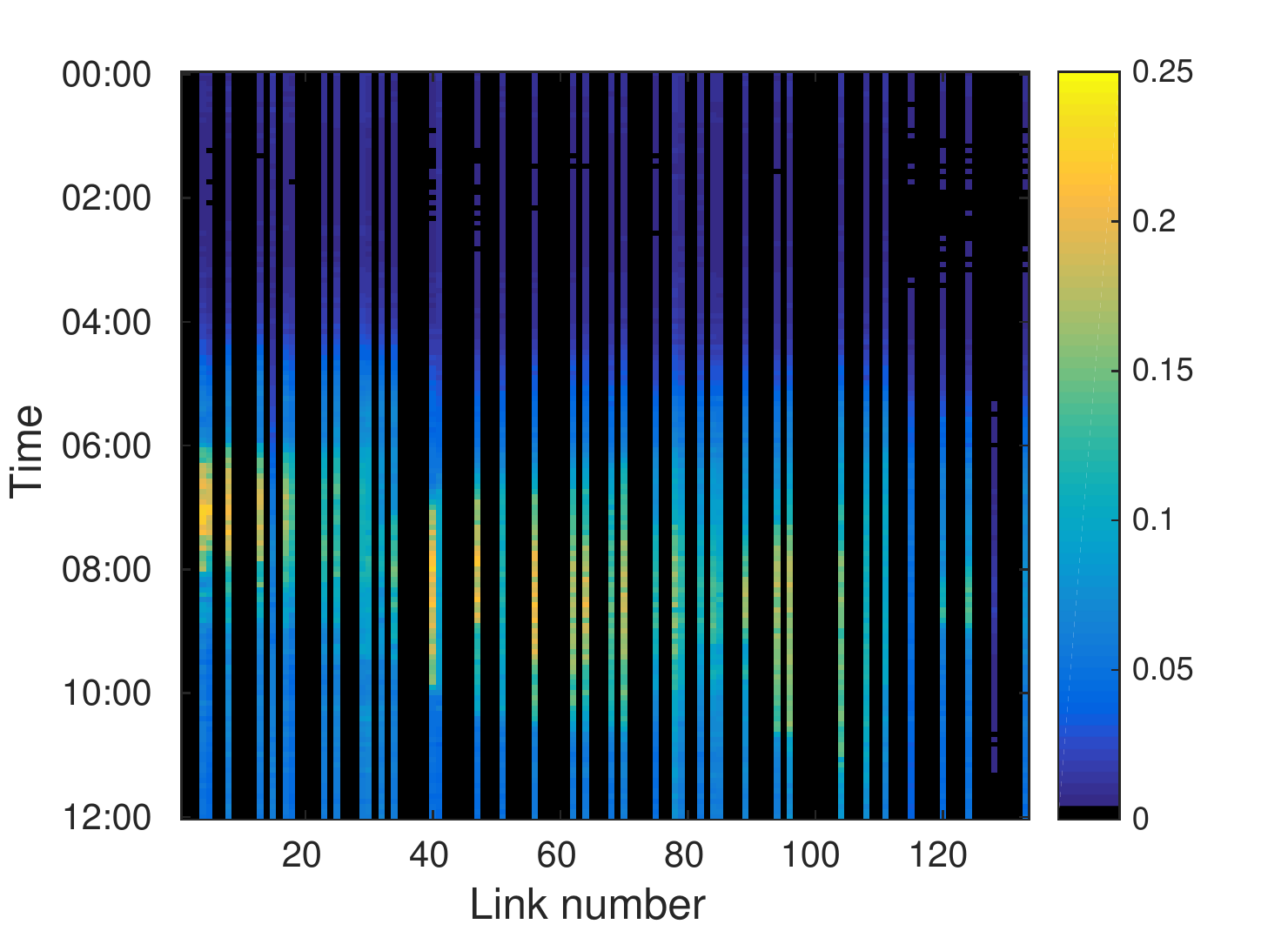}
    \label{fig:Oct22Loops}} 
  \caption{PeMS loop data of the site (veh/m). Black = no data. (a) Oct 13, 2014 (calibration date). (b) Oct 22, 2014 (test date). Note that the traffic patterns exhibit broader congestion periods on Oct 22.}
  \label{fig:realData}
\end{figure*}

Fig.~\ref{fig:realData} presents the raw loop data used in this procedure. Note that the traffic behavior on October 22 was different from that of October 13, with much larger and longer-lasting areas of congestion (high density).

Of the 35 working detectors on this date, 15 were randomly selected as a ``test'' set whose density measurements were compared against the estimates, and the measurements of the remaining detectors were provided to the filters.  In the CTM simulations, onramps with working detectors had their measured flow input to the simulation on a five-minute delay with a simple zero-order-hold assumption (that is, that the average flow for a given five-minutes would be equal to the average measured flow of the previous five minutes), with per-timestep additive Gaussian noise sampled from a Gaussian distribution with standard deviation equal to 15\% of this nominal value. All other simulation parameters were set up identically to the simulated experiment.

A problem common to attempts to use third-party probe data for traffic estimation is matching noisy GPS points to individual roads, and filtering out those points that are not sent from vehicles in transit in the direction of interest. Map-matching schemes attempt to perform this filtering by reconstructing a vehicle's trajectory along the network and matching individual reports to the mapped links. In the present experiment, though, our network is very simple; we only seek to estimate the state along the mainline, so we used a simpler probe point filtering scheme: non-overlapping rectangular bounding boxes were drawn around each link and probe points were assigned to the link whose bounding box they fell within, if any. Further, probe points with reported headings outside of $15^\circ$ of a link's end-to-end bearing were discarded. To filter out erroneous data, individual measurements $y_i$ that evaluated to a likelihood $g( y_i | x_{i,p} ) \approx 0$ for all particles $p$ were excluded from the calculations. Such measurements may be the result of, e.g., parked cars within the bounding box or faulty equipment.

The probe data were tagged with a hashed device identifier; after our crude map-matching scheme, probe points associated with 2613 unique devices remained. Over this same 12-hour period, the loop detectors along the freeway reported an average cumulative flow of approximately 182,000 vehicles along the mainline, resulting in an estimated penetration rate of roughly 1.42\% of our probe data.

\begin{figure*}
  \centering
  \subfloat[]{
    \includegraphics[width=20pc,trim=0 0 2pc 1.5pc,clip]{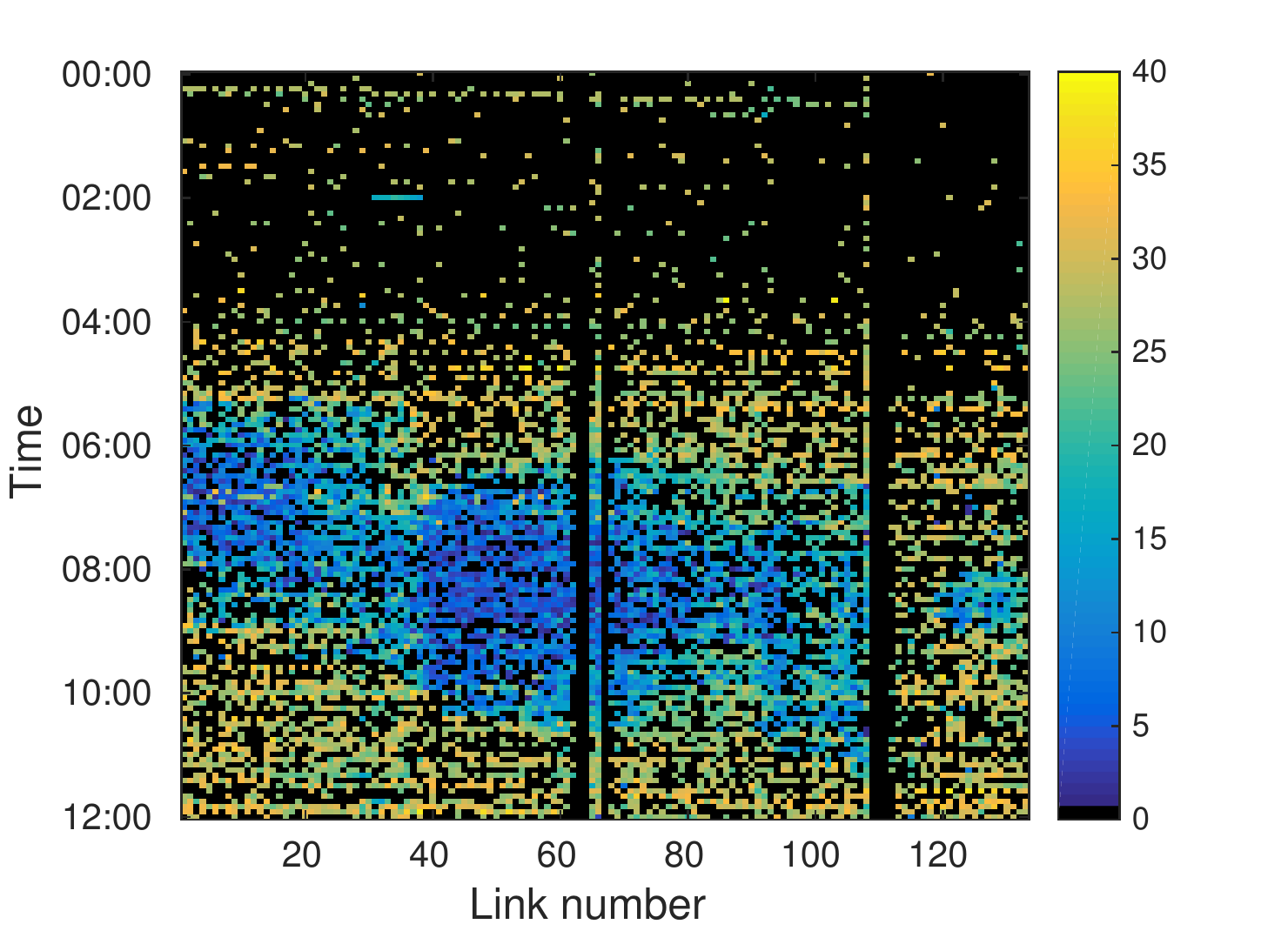}
    \label{fig:Oct22Probes}} 
  \subfloat[]{
    \includegraphics[width=20pc,trim=0 0 2pc 1.5pc,clip]{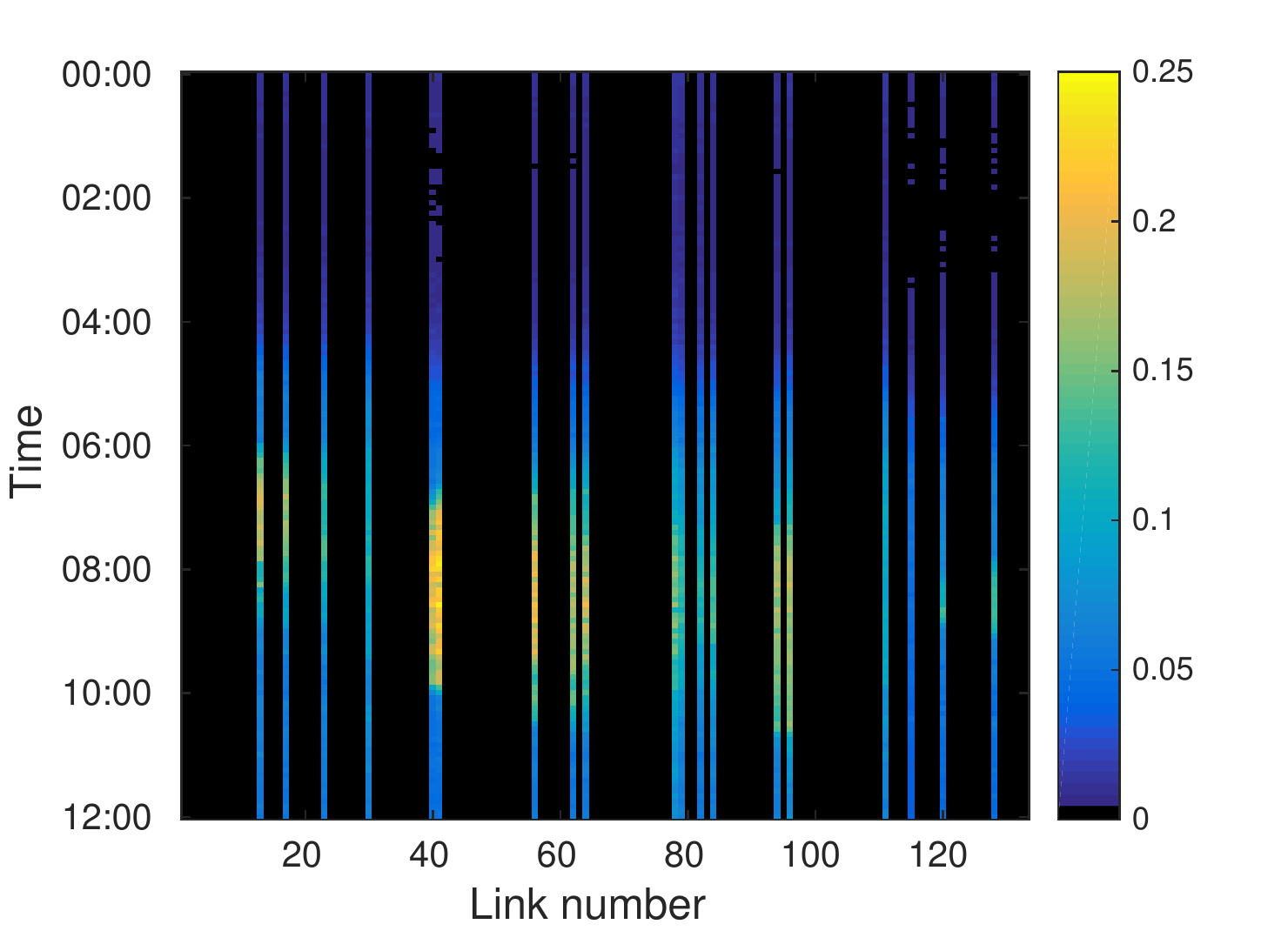}
    \label{fig:Oct22LoopSubset}} \\
  \subfloat[]{
    \includegraphics[width=20pc,trim=0 0 2pc 1.5pc,clip]{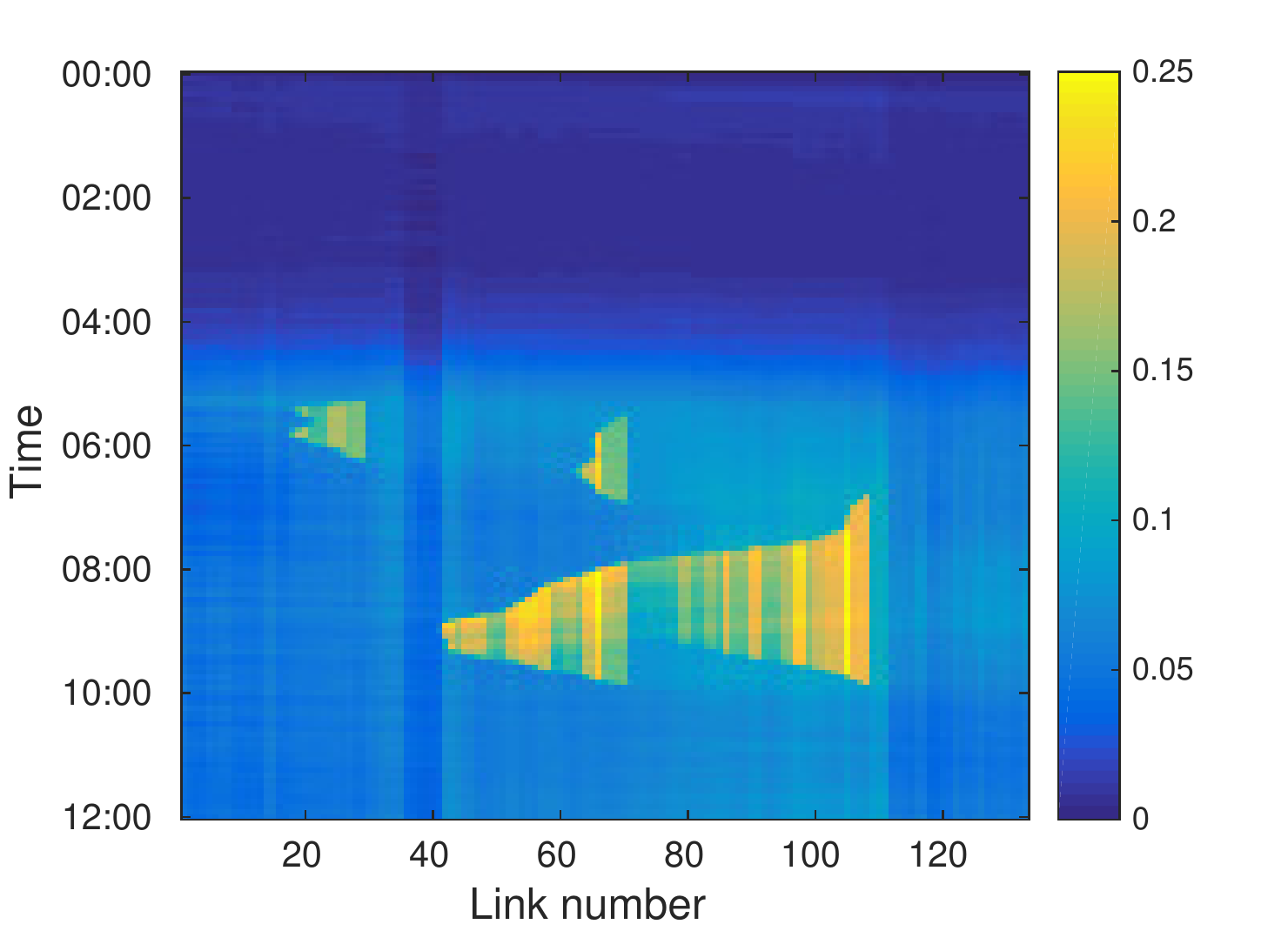}
    \label{fig:Oct22BestCase}} 
  \subfloat[]{
    \includegraphics[width=20pc,trim=0 0 2pc 1.5pc,clip]{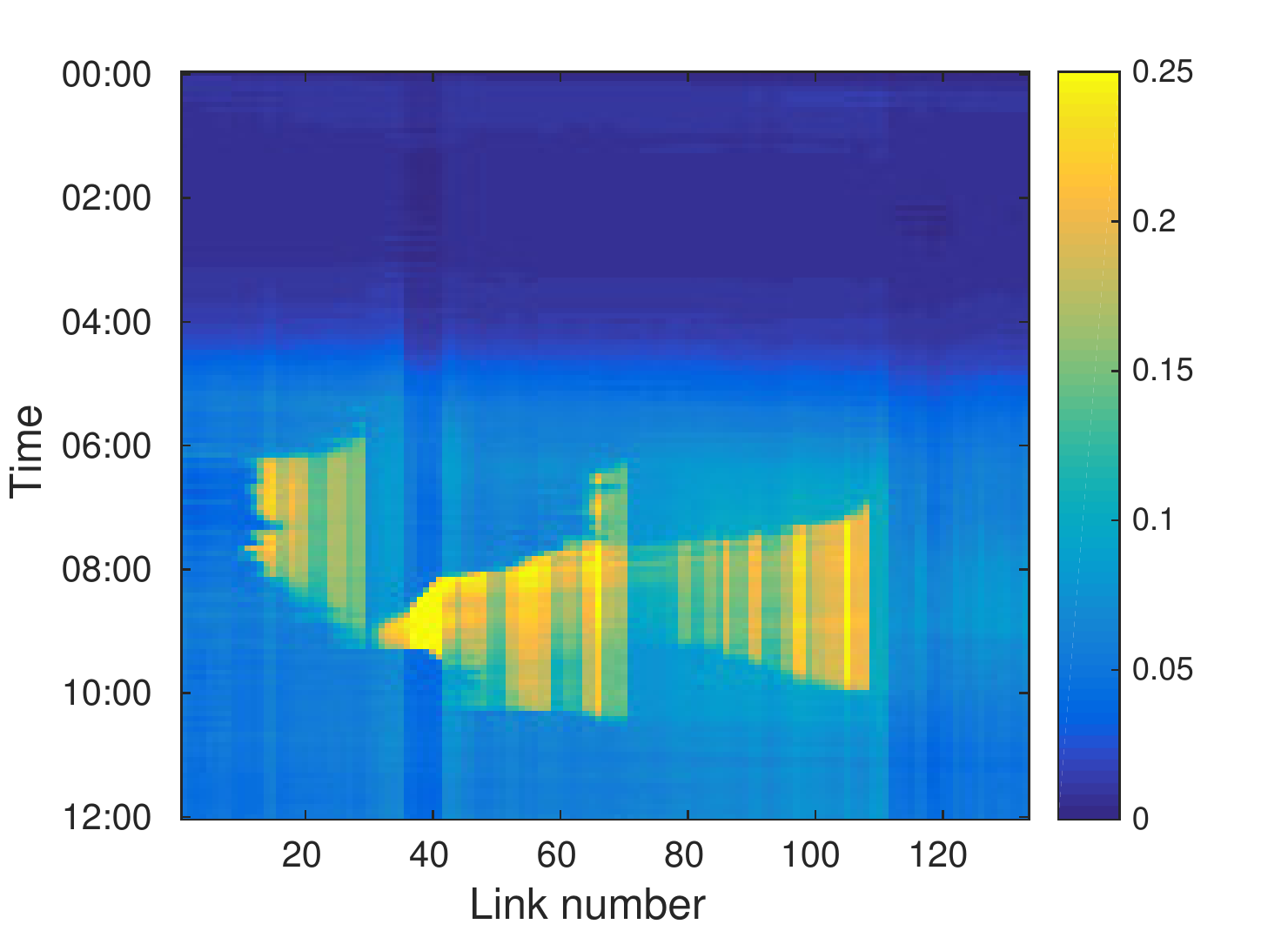}
    \label{fig:Oct22EstimatedWithLoops}} \\
  \subfloat[]{
    \includegraphics[width=20pc,trim=0 0 2pc 1.5pc,clip]{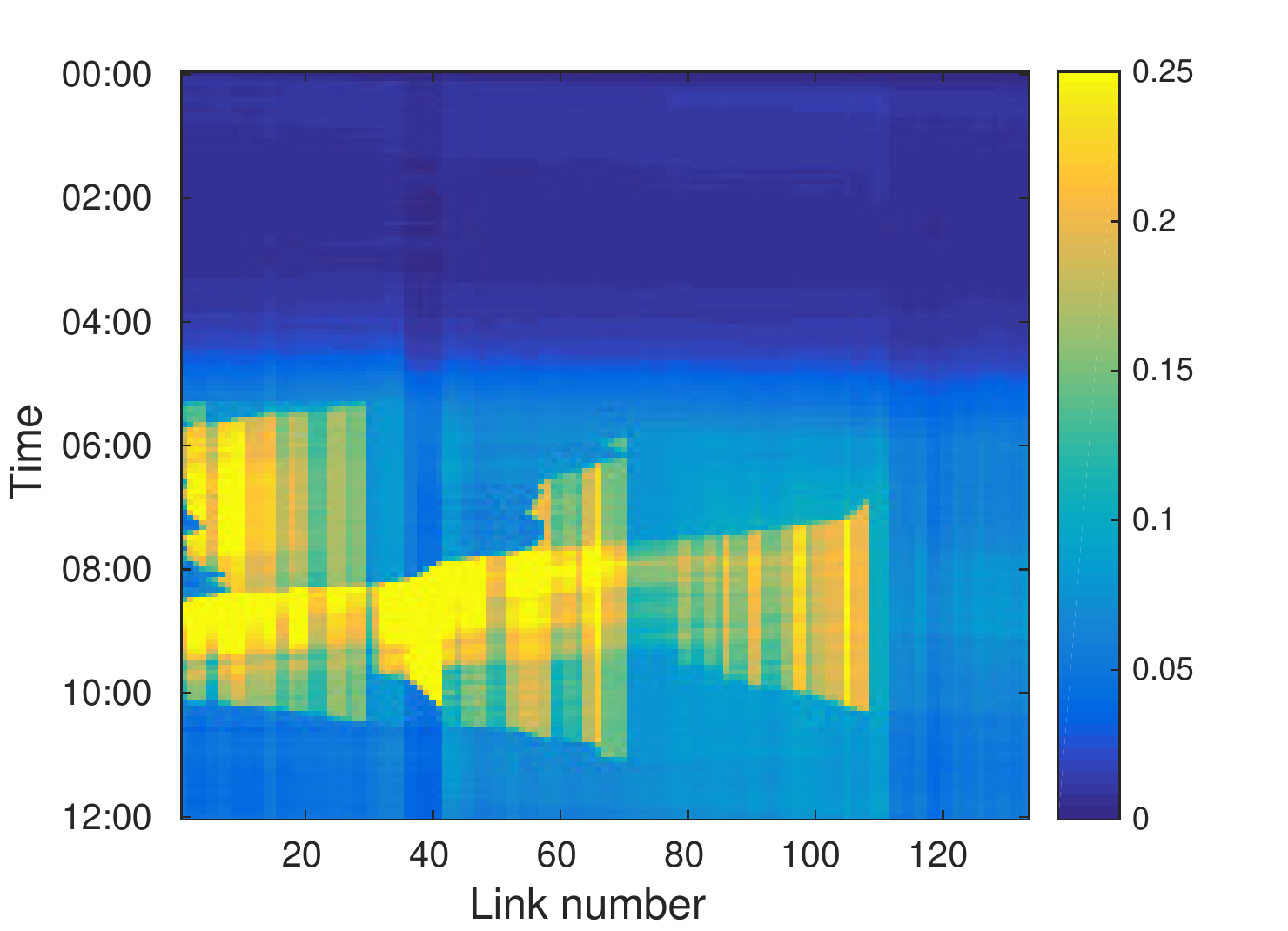}
    \label{fig:Oct22EstimatedWithProbes}} 
  \subfloat[]{
    \includegraphics[width=20pc,trim=0 0 2pc 1.5pc,clip]{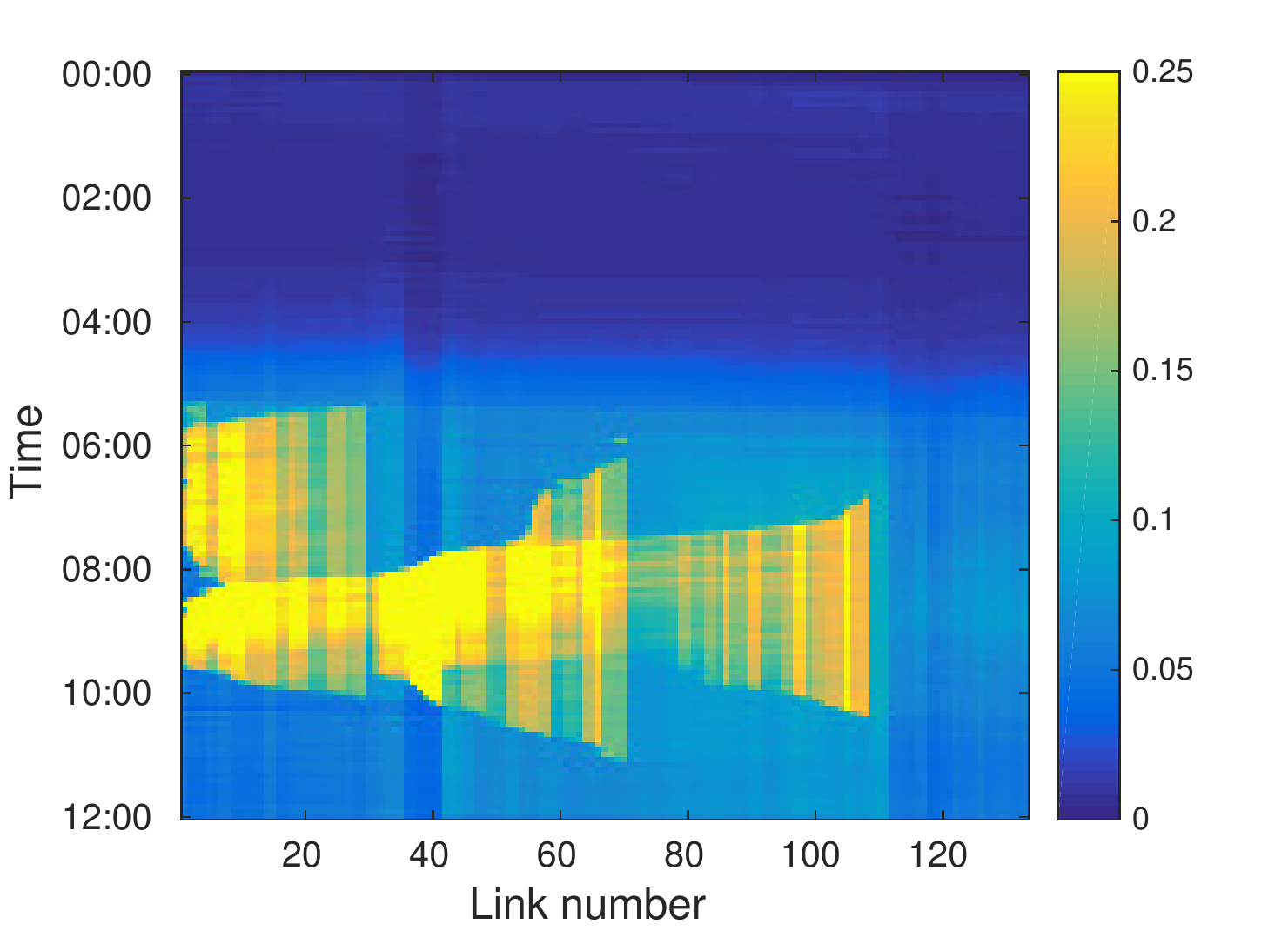}
    \label{fig:Oct22EstimatedWithBoth}}
  \caption{Density estimation of traffic density of the corridor on Oct 22, 2014 with real data. Black = no data. (a) Raw probe speeds (m/s). (b) Loop density subset provided to filters (veh/m). (c) ``Best-case'' open-loop prediction (veh/m). (d) Estimation with loop data only (veh/m). (e) Estimation with probe data only (veh/m). (f) Estimation with data fusion (veh/m).}
  \label{fig:realDataResults}
\end{figure*}

Results are shown in Fig.~\ref{fig:realDataResults}. It is known that traffic models are limited in open-loop (that is, without data filtering) prediction accuracy due to factors including day-to-day variability in driver and road characteristics and a low signal-to-noise ratio exhibited in freeway offramp and onramp measurements (i.e., PDE boundary conditions); Fig.~\ref{fig:Oct22BestCase} shows a baseline simulation using as-detected on- and off-ramp flow values from Oct 22, but no mainline measurements, to demonstrate a base level of error to which filtered estimates may be compared against. Qualitatively, one sees that the use of the probe data (shown in Fig.~\ref{fig:Oct22Probes}) produces a density estimate with larger and longer-lasting congestion periods (compare Figs. \ref{fig:Oct22EstimatedWithLoops}, \ref{fig:Oct22EstimatedWithBoth}). As a specific example, we wish to direct the reader's attention to the congestion patterns in the first ~10 links. Note that in Fig.~\ref{fig:Oct22Loops}, a period of high density occurs in the first few detectors during 6-8 AM. These detectors are excluded from the simulation~(Fig. \ref{fig:Oct22LoopSubset}), and accordingly the density-only particle filter does not predict that congestion occurs in these links during this period (Fig.~\ref{fig:Oct22EstimatedWithLoops}). There exist probe measurements during that period that show low velocity (Fig.~\ref{fig:Oct22Probes}), and when the PF is run with these measurements available, congestion is predicted (Fig.~\ref{fig:Oct22EstimatedWithProbes}).

In fact, our algorithm as implemented for this test may be a little \emph{too} eager to assimilate probe measurements. Note that after 10 AM on the same few detectors, the congestion wholly clears in reality (Fig.~\ref{fig:Oct22Loops}). However, the RBPF still sees a number of probe data that continue to report low velocity (observe the cloud of blueish points in this area in Fig. ~\ref{fig:Oct22Probes}). With no loop data to counteract this, the RBPF continues to estimate congestion until the cloud of low-velocity probe data clears (Fig.~\ref{fig:Oct22EstimatedWithBoth}). Recall that in our construction of the RBPF likelihood functions discussed at the beginning of this section, the probe measurements are taken to be distributed symmetrically about the CTM-predicted velocity $\bar{x}^v_t$ with exponentially-decaying outliers (i.e. tail behavior) due to the Gaussian assumption. However, the empirical evidence here shows that, at least in this location, this is not a good assumption. Instead, a large number of probe data reported slow velocities despite the loop detectors in the area reporting low density. The RBPF then estimated a region of high density that would be in agreement with high density that would be in agreement with the low-velocity probe data through~\eqref{eq:velocity}. This suggests the need for investigation into the tail behavior and non-stationarity (in the timeseries sense) of the PDF $p_\theta( y^v_{t, L(i)} | \bar{x}^v_{t, L(i)})$.

Fusing the loop and probe data (Fig.~\ref{fig:Oct22EstimatedWithBoth}) produces an estimate that appears largely similar to the probe-only estimate, but tends to obtain lower error (Table~\ref{tab:realMAPE}). Data fusion generally outperformed use of the disjoint sets in estimating the density measurements for most detectors in terms of MAPE metric, despite oversensitivity to probe data resulting in errors.

\begin{table*}
\renewcommand{\arraystretch}{1.2}
\caption{Density MAPE Metrics for Real-Data Experiment}
\label{tab:realMAPE}
  \centering
  \begin{tabular}{ l | l l l l }
  & \multicolumn{4}{c}{MAPE} \\
  \cline{2-5}
  PeMS detector ID & Open-loop simulation & Loop data only & Probe data only & Fused data \\
  \hline
  774012 & 39.20\% & 28.38\% & 32.76\% & 27.20\% \\
  717599 & 39.32 & 27.54 & 31.77 & 27.00 \\
  717634 & 38.76 & 36.19 & 34.70 & 32.87 \\
  718210 & 38.93 & 35.17 & 28.39 & 25.62 \\
  717664 & 46.24 & 35.29 & 37.40 & 31.73 \\
  717669 & 32.56 & 19.75 & 22.53 & 11.43 \\
  764146 & 44.08 & 35.95 & 26.82 & 21.89 \\
  717649 & 18.08 & 14.72 & 15.29 & 11.76 \\
  717653 & 26.85 & 22.37 & 21.05 & 16.48 \\
  717637 & 24.46 & 25.70 & 24.03 & 25.81 \\
  717675 & 38.32 & 53.57 & 35.84 & 32.21 \\
  772918 & 32.05 & 56.80 & 33.20 & 30.16 \\
  717688 & 29.94 & 59.74 & 31.29 & 27.49 \\
  761374 & 52.94 & 44.44 & 40.75 & 35.63 \\
  772888 & 30.89 & 47.14 & 29.19 & 18.49 \\
  \hline
  Mean$\pm \sigma$ & $35.51\pm8.96$ & $36.18\pm13.73$ & $29.67\pm6.81$ & $25.05\pm7.55$
  \end{tabular}
\end{table*}

One may notice the high amounts of error in these results compared to the simulated results in Table 1. This is due to two factors. First, the CTM is a relatively simple model of traffic, and while it can easily capture broad congestion patterns, it cannot reproduce some higher-order traffic phenomena to high accuracy. Second, the “ground truth” density measurements the estimates are compared against are not actually the true density values, but rather the noisy measurements from the loop detectors. This type of loop detectors is known to be noisy \cite{chen2003errors}. We did not attempt to denoise the loop measurements. A low MAPE value would thus require that the model estimate reproduce the loop's sensor noise, which we feel is infeasible.

\section{Conclusions and future work}
\label{sec:conclusion}
This paper introduced a filtering scheme for tractably estimating vehicle density while assimilating probe velocity data in the structure of a Rao-Blackwellized particle filter by exploiting some conditional independence assumptions. Use of these assumptions led to a model that implicitly estimated vehicle velocity for the purposes of filtering, which we referred to as a pseudo-second-order model. We demonstrated the effectiveness of our model for a long freeway corridor.

Our numerical experiments were based on a freeway, but the particle filter itself is not restricted to traffic networks of this rigid structure, and is applicable to more complex networks. Of particular interest might be interconnected urban networks; these networks typically have lower fixed detection infrastructure then freeways, and our second experimental result of solely velocity data being used to estimate density is encouraging for this application. Any application, though, would have to overcome difficulties in accurately matching probe measurements to individual road segments, which is itself a difficult problem.

Immediate theoretical avenues for investigation that present themselves are estimation of the PDFs $p_\theta( \bar{x} | x^\rho)$ and $p_\theta( y^v | \bar{x}^v)$ in~\eqref{eq:firstOrderLikelihoodFactoring} (denoting the distribution of a road segment's average velocity about a model-predicted velocity, and the distribution of individual vehicle measurements about a link's average velocity, respectively). Typically these PDFs have taken assumed form~\cite{sumalee2011stochastic}, but estimation of these distributions from data would allow for more accurate filtering models. Another item of immediate applicability is the use of a particle smoother to recreate the state trajectory while taking into account future probe measurements, but given the high dimensionality of traffic network models, this would not be a straightforward application either.

\section*{Acknowledgement}
The authors thank Gabriel Gomes, who provided suggestions in response to an earlier version of this paper and shared his expertise on traffic modeling. M. Wright also wishes to thank David Shulman for previous collaboration related to this problem.

\iftoggle{eprint}{
\bibliographystyle{abbrv}
\bibliography{bibliography}}{
\bibliographystyle{IEEEtran}
\bibliography{IEEEabrv,bibliography}
}

\end{document}